\documentclass[iop,apj,onecolumn]{emulateapj}

\usepackage{graphicx}
\usepackage{amssymb}
\usepackage{amsmath}
\usepackage{multirow}
\usepackage[linktocpage=true]{hyperref}

\shorttitle{Disc perturbation of a Schwarzschild black hole}
\shortauthors{P. \v{C}\'{\i}\v{z}ek and O. Semer\'ak}

\begin{document}

\title{Perturbation of a Schwarzschild black hole due to a rotating thin disc}

\author{P. \v{C}\'{\i}\v{z}ek and O. Semer\'ak}
\affil{Institute of Theoretical Physics,
       Faculty of Mathematics and Physics,
       Charles University, Prague, Czech Republic}
\email{oldrich.semerak@mff.cuni.cz}

\begin{abstract}
\cite{bib:Will1974} treated the perturbation of a Schwarzschild black hole due to a slowly rotating light concentric thin ring by solving the perturbation equations in terms of a multipole expansion of the mass-and-rotation perturbation series. In the Schwarzschild background, his approach can be generalized to the perturbation by a thin disc (which is more relevant astrophysically), but, due to a rather bad convergence properties, the resulting expansions are not suitable for specific (numerical) computations. However, we show that Green's functions represented by the Will's result can be expressed in a closed form (without multipole expansion) which is more useful. In particular, they can be integrated out over the source (thin disc in our case), to yield well converging series both for the gravitational potential and for the dragging angular velocity. The procedure is demonstrated, in the first perturbation order, on the simplest case of a constant-density disc, including physical interpretation of the results in terms of a one-component perfect fluid or a two-component dust on circular orbits about the central black hole. Free parameters are chosen in such a way that the resulting black hole has zero angular momentum but non-zero angular velocity, being just carried along by the dragging effect of the disc.
\end{abstract}


\section{Introduction}

Disc-like structures around very compact bodies are likely to play a key role in the most energetic astrophysical sources like active galactic nuclei, X-ray binaries, supernovas and gamma-ray bursts. Analytical modeling of such structures relies on various simplifying assumptions, the basic ones being their stationarity, axial symmetry and test (non-gravitating) nature (see e.g. \citealt{bib:Kato2008}). The last assumption is justified by two arguments: (i) in the above astrophysical systems, the disc mass is typically much smaller than that of the black hole (or neutron star) in their center, so the latter surely dominate the gravitational potential as well as the ``radial'' field; (ii) black holes are the strongest possible (extended) gravitational sources (and neutron stars are just slightly less compact), so they would —- at least in a certain region —- even dominate the field if their mass was less than that of the matter around. However, such arguments need not hold for a latitudinal component of the field (namely perpendicular to the disc),\footnote
{For a Schwarzschild black hole which is spherically symmetric, there is no latitudinal field of course, so {\em any} additional source would automatically dominate this component.}
and, most importantly, the additional matter may in fact dominate the second and higher derivatives of the metric (curvature). These higher derivatives are in turn crucial for {\em stability} of the matter's motion, and thus the tricky issue of self-gravity enters the problem. Actually, a real, massive matter may thus assume quite a different configuration than a test matter \citep{bib:Abr1984}. One should also add that even if the accreting matter really had only a tiny effect on the geometry, it could still change the observational record of the source significantly, in particular, it may change the long-term dynamics of bodies orbiting in the system (e.g. \citealt{bib:SukovaSem2013} and references therein).

Hence, the properties of accretion systems may be sensitive to the precise shape of the field \citep{bib:Semer2003,bib:Semer2004}. Unfortunately, general relativity is non-linear and the fields of multiple sources mostly cannot be obtained by a simple superposition. Such more complicated fields are being successfully treated numerically (this even applies to strongly time-dependent cases including gravitational collapse, collisions and waves), but, for the present, the compass of explicit analytical solution terminates at systems with very high degree of symmetry, practically at static and axially symmetric cases. It would be most desirable to extend this to {\em stationary} cases, namely those admitting rotation. Stationary axisymmetric problem is usually represented in the form of the Ernst equation, but actually being tackled is the corresponding linear problem (Lax pair of equations whose integrability condition is just the Ernst equation). {\em Exact} solutions of this problem have been searched for in several ways. \cite{bib:KleinRich2005} and \cite{bib:Meinel2008} summarized ``straightforward'' but rather involved treatment of the respective boundary-value problem, providing both the black-hole and the thin-disc solutions, plus prospects of how to also obtain their non-linear superpositions. Other attempts employed the ``solution-generating'' techniques —- mathematical procedures which transform one stationary axisymmetric metric into another and can in principle provide any solution of this type. The practical power of these methods strongly depends on how simple is the ``seed'' metric, so usually a {\em static} one is started from. Using the soliton (inverse-scattering) method of Belinsky \& Zakharov, \cite{bib:Tomi1984}, \cite{bib:KroriBat1990}, \cite{bib:ChauDas1997} and \cite{bib:ZellSem2000} generated black holes immersed in external fields, but at least the case corresponding to a hole surrounded by a thin disc (\citealt{bib:ZellSem2000}) turned out to be unphysical (\citealt{bib:Semer2002}). More successful seem to have been \cite{bib:Breton1997} who started from a different representation of the static axisymmetric seed and managed to ``install'' a rotating black hole in it (see also \cite{bib:Breton1998} for a charged generalization).

If the external-matter gravity is weak, the problem may be treated as a small {\em perturbation} of the central-source metric, determined by {\em linearized} Einstein equations. In doing so, one can restrict to a special type of perturbations, for example, to stationary and axisymmetric ones. The method can be iterated; in a limit case (many iterations), it goes over to a solution in terms of series. The result then need not any longer represent a tiny variation of any ``almost right'' metric: it may even be put together on Minkowski background, with the ``strong'' part (e.g. a black hole) ``dissolved'' within the fundamental systems of the equations. The main problem of this scheme is convergence and meaning of the series.

No less than 43 years ago, the paper by \cite{bib:Will1974}, published in this journal, became a seminal reference in the subject. It provided the gravitational field of a light and slowly rotating thin equatorial ring around an (originally Schwarzschild) black hole by mass-and-rotational perturbation of the Schwarzschild metric. (See also the following paper \cite{bib:Will1975} where basic properties of the obtained solution were discussed.) Unfortunately, the perturbation-scheme success depends strongly on how simple the background metric is -- and the Schwarzschild metric is exceptionally simple: the Will's procedure cannot be simply extended to a Kerr background. Consequently, only partial questions have been answered explicitly in these directions, in particular the one of deformation of the Kerr black-hole horizon \citep{bib:Demianski1976,bib:Chrzanowski1976} (interestingly, these two results do not agree on certain points, mainly in the limit of an extreme black hole).\footnote
{Note that another approximation possibility is the post-Newtonian expansion. The composition of a rotating gravitational center with a massive ring in Keplerian rotation was tackled, using the gravito-electromagnetic analogy, by \cite{bib:Ruggiero2016}.}

Recently the Will's black-hole--ring problem has been revisited by \cite{bib:SanoTagoshi2014}, but using the perturbation approach of Chrzanowski, Cohen and Kegeles in which the metric is found on the basis of solution of the Teukolsky equation for the Weyl scalars.
The Will's results have also been followed by Hod who analyzed the behaviour of the innermost stable circular orbit in the black-hole--ring field \citep{bib:Hod2014} and the relation between the angular velocity of the horizon and the black-hole and ring angular momenta \citep{bib:Hod2015}.

In the present note, we check whether the Will's scheme can be adapted to the case of a thin equatorial stationary and axisymmetric disc. [Preliminary results were presented in \cite{bib:CizSem2009} and \cite{bib:Ciz2011}.] In converging to a positive answer, we first observed that the expansions in spherical harmonics that typically come out in this approach (even in computing just the linear terms) converge rather badly and their numerical processing is problematic. Much more effective is the usage of Green functions of the problem, namely the perturbations generated by an infinitesimal ring. We have been able to express the Green functions in such a manner that the linear perturbation due to a thin disc can be obtained in a closed form.

The paper is organized as follows. In section \ref{ch:EinsteinEQ} we introduce equations describing the gravitational field of a thin disc. Section \ref{ch:Perturbation} summarizes Will's approach and section \ref{ch:Convergence} discusses its (un)suitability for a numerical treatment. In section \ref{ch:GF} we compute Green functions of the problem in a closed form and in section \ref{ch:Disc} we show, on {\em linear} perturbation by a thin annular concentric disc, that they can be integrated in order to obtain a perturbation generated by a given (stationary and axisymmetric) distribution of mass. The resulting series converge much better and allow to compute specific configurations explicitely.

Notation and conventions: our metric signature is ($-$$+$$+$$+$) and geometrized units are used in which $c\!=\!G\!=\!1$; Greek indices run 0-3 and partial derivative is denoted by a comma. Complete elliptic integrals are given in terms of the {\em modulus} $k$, so
\[K(k):=\int_0^{\frac{\pi}{2}}\!\!\!\frac{{\rm d}\alpha}{\sqrt{1-k^2\sin^2\alpha}} \;,
  \qquad
  E(k):=\int_0^{\frac{\pi}{2}}\!\!\sqrt{1-k^2\sin^2\alpha}\;{\rm d}\alpha \;,
  \qquad
  \Pi(n,k):=\int_0^{\frac{\pi}{2}}\!\!\!\frac{{\rm d}\alpha}{(1-n\,\sin^2\alpha)\sqrt{1-k^2\sin^2\alpha}} \;.\]

\section{Black-hole \& thin-disc system: Einstein equations and boundary conditions}
\label{ch:EinsteinEQ}

We will search for the black-hole--disc field by perturbation of the Schwarzschild metric, while restricting to the simplest space-times which can host {\em rotating} sources, namely to those which are stationary and axially symmetric. In addition, we will consider asymptotically flat space-times, without cosmological term, and will require their orthogonal transitivity (i.e., the motion of sources will be limited to stationary circular orbits). In such space-times, the time and axial Killing vector fields $\eta^\mu\!=\!\frac{\partial x^\mu}{\partial t}$ and $\xi^\mu\!=\!\frac{\partial x^\mu}{\partial \phi}$ exist and commute, and the tangent planes to meridional directions (locally orthogonal to both Killing vectors) are integrable. Needless to say, it is assumed that there exists an axis of the space-like symmetry, namely a connected 2D (time-like) set of fixed points of the space-like isometry. In isotropic-type spheroidal coordinates ($t$,$r$,$\theta$,$\phi$) (of which $t$ and $\phi$ have been chosen as parameters of the Killing symmetries), the metric with these properties can -- for instance -- be written in the ``Carter-Thorne-Bardeen'' form (e.g. \citealt{bib:Bardeen1973})
\begin{equation}
  {\rm d}s^2 = -e^{2\nu}{\rm d}t^2+B^2 r^2 e^{-2\nu}\sin^2\theta\;({\rm d}\phi-\omega{\rm d}t)^2
               +e^{2\zeta-2\nu}({\rm d}r^2+r^2{\rm d}\theta^2) \,,
\end{equation}
where the unknown functions $\nu$, $B$, $\omega$ and $\zeta$ depend only on $r$ and $\theta$ covering the meridional surfaces. Besides the above coordinates, we will also occasionally use the Weyl-type cylindrical coordinates $\rho\!=\!r\sin\theta$ and $z\!=\!r\cos\theta$. 

Apart from the asymptotic flatness, the boundary conditions have to be fixed on the symmetry axis, on the black-hole horizon and on the external-source surface. Regularity of the {\em axis} (local flatness of the orthogonal surfaces $z\!=\!{\rm const}$ at $\rho\!=\!0$) requires that $e^\zeta\!\rightarrow\!B$ at $\rho\!\rightarrow\!0^+$. The invariants
$g_{tt}\!=\!g_{\alpha\beta}\eta^\alpha\eta^\beta$,
$g_{t\phi}\!=\!g_{\alpha\beta} \eta^\alpha \xi^\beta$ and
$g_{\phi\phi}\!=\!g_{\alpha\beta} \xi^\alpha \xi^\beta$
have to be even functions of $\rho$ (in order not to induce a conical singularity on the axis). Should the circumferential radius $\sqrt{g_{\phi\phi}}$ grow linearly with proper cylindrical radius
$\rho\,[e^{\zeta-\nu}]_{\rho=0}$, thus with $\rho$, there must be $g_{\phi\phi}\!\approx\!O(\rho^2)$, and, demanding the finiteness of $\omega$, also $-g_{t\phi}\!=\!g_{\phi\phi}\omega\!\approx\!O(\rho^2)$.

The stationary {\em horizon} is characterized by $e^{2\nu}\!=\!0$ and $\omega\!=\!{\rm const}\!=:\!\omega_{\rm H}$ (in our coordinates it specifically means that $\omega_{,\theta}\!=\!0$ there). In order that the azimuthal and latitudinal circumferences of the horizon be positive and finite, the functions $Bre^{-2\nu}$ and $e^{2\zeta-2\nu}$ have to be such; the latter ensures regularity of $g_{rr}$ as well. Hence, $Br\!=\!0$ and $e^{2\zeta}\!=\!0$ on the horizon. (Let us add in advance that the field equations also imply that $\omega_{,r}/B$ and $\nu_{,r}\omega_{,r}$ have to be finite on the horizon, so $\omega_{,r}$ has to vanish there as well.)

Now for boundary conditions on the external source. We assume that the latter has the form of an {\em infinitesimally thin disc} in the equatorial plane $z\!=\!0$, stretching over some interval of radii lying above the central black-hole horizon. We assume that the disc bears neither charge nor current (there are no EM fields) and that the space-time is reflection symmetric with respect to its plane. The metric is then continuous everywhere, but has finite jumps in the first normal derivatives $g_{\alpha\beta,z}$ across the disc. The functions $\nu$, $B$, $\omega$ and $\zeta$ must be even in $z$, their $z$-derivatives are odd in $z$, and even powers and multiples of derivatives (for example, $B_{,z}\nu_{,z}$) are even in $z$ (therefore they do not jump across $z\!=\!0$).

In order that the space-time be stationary, axially symmetric and orthogonally transitive, the disc elements must only move along surfaces spanned by the Killing fields, namely they must follow spatially circular orbits with steady angular velocity $\Omega\!=\!\frac{{\rm d}\phi}{{\rm d}t}$. This corresponds to four-velocity
\begin{equation}
  u^\alpha = \frac{\eta^\alpha+\Omega\xi^\alpha}{\left|\eta^\alpha+\Omega\xi^\alpha\right|}
           = u^t (1,0,0,\Omega) \,, \qquad
  u_\alpha = -u^t e^{2\nu}\delta^t_\alpha + u^t B \rho v\,(-\omega,0,0,1)
\end{equation}
with
\[(u^t)^2 =\frac{e^{-2\nu}}{1-B^2\rho^2 e^{-4\nu}(\Omega-\omega)^2}
          =\frac{e^{-2\nu}}{1-v^2} \;,
  \qquad\quad {\rm where} \qquad
  v:=B\rho e^{-2\nu}(\Omega-\omega)=\sqrt{g_{\phi\phi}}\,e^{-\nu}(\Omega-\omega)\]
represents linear velocity with respect to the local zero-angular-momentum observer (ZAMO).
For the thin discs ($T^z_z\!=\!0$, $T^\rho_z\!=\!0$) without radial pressure ($T^\rho_\rho\!=\!0$), the surface energy-momentum tensor
\begin{equation}
  \int\limits_{-\infty}^{\infty} T^\alpha_\beta g_{zz}\,{\rm d}z
  =\int\limits_{z=0^-}^{z=0^+} T^\alpha_\beta e^{2\zeta-2\nu}{\rm d}z
  =: S^\alpha_\beta(\rho) 
  \qquad\Longleftrightarrow\qquad
  T^\alpha_\beta e^{2\zeta-2\nu} =: S^\alpha_\beta(\rho) \delta(z)
\end{equation}
has only three non-zero components ($S^t_t$,$S^t_\phi$,$S^\phi_\phi$), representing energy density, orbital-momentum density and azimuthal pressure, respectively. If $(S^\phi_\phi-S^t_t)^2+4S^t_\phi S^\phi_t \geq 0$, it can be diagonalized to
$S^{\alpha\beta}=\sigma u^\alpha u^\beta+Pw^\alpha w^\beta$,
where $\sigma$ and $P$ (more precisely, $\sigma e^{\nu-\zeta}$ and $Pe^{\nu-\zeta}$) stand for the surface density and azimuthal pressure in a co-moving frame and $w^\alpha$ is the ``azimuthal'' vector perpendicular to $u^\alpha$, with components $w^\alpha\!=\!\frac{1}{\rho B}\,(u_\phi,0,0,-u_t)$, $w_\alpha\!=\!\rho B\,(-u^\phi,0,0,u^t)$. Hence, the surface-tensor components read
\begin{equation}  \label{S-components}
  S^t_t = -\sigma -(\sigma+P)\,u^\phi u_\phi \,, \qquad
  S^t_\phi = (\sigma + P)\,u^t u_\phi \,, \qquad
  S^\phi_\phi = P + (\sigma+P)\,u^\phi u_\phi \,.
\end{equation}

Orthogonally transitive stationary and axisymmetric space-times are described by 5 independent Einstein equations. In our case of a thin disc, the energy-momentum tensor has only $T^t_t$, $T^t_\phi$ and $T^\phi_\phi$ components and the equations read\footnote
{Equation (\ref{eq:EFE+}) is {\em not} independent, but we include it here since it provides the jump of $\zeta_{,z}$ across the equatorial plane given later in section \ref{jumps}.}
\begin{align}
  & \nabla\!\cdot\!(\rho\nabla B) = 0 \,, \label{eq:EFE1} \\
  & \nabla\!\cdot\!(B\nabla\nu)-\frac{B^3\rho^2}{ 2e^{4\nu}}(\nabla\omega)^2
       = 4\pi B e^{2\zeta-2\nu}\left(T^\phi_\phi-2\omega T^t_\phi-T^t_t\right)
       = 4\pi B\,(\sigma+P)\,\frac{1+v^2}{1-v^2}\,\delta(z) \,, \label{eq:EFE2} \\
  & \nabla\!\cdot\!(B^3\rho^2e^{-4\nu}\nabla\omega)
       = -16\pi B e^{2\zeta-2\nu} T^t_\phi
       = -16 \pi B^2\rho e^{-2\nu}(\sigma+P)\,\frac{v}{1-v^2}\,\delta(z) \,, \label{eq:EFE3} \\
  & \zeta_{,\rho\rho}+\zeta_{,zz}
         +(\nu_{,\rho})^2+(\nu_{,z})^2
         -\frac{3B^2\rho^2}{4e^{4\nu}}\left[(\omega_{,\rho})^2+(\omega_{,z})^2\right]
       = 8\pi e^{2\zeta-2\nu}\left(T^\phi_\phi-\omega T^t_\phi\right)
       = 8\pi\,\frac{\sigma v^2+P}{1-v^2}\,\delta(z) \,,  \label{eq:EFE+} \\
  & \zeta_{,\rho}(B\rho)_{,\rho}-\zeta_{,z}(B\rho)_{,z}
       = -B\rho\left[(\nu_{,\rho})^2-(\nu_{,z})^2\right]
         -\frac{1}{2}\left[(B\rho)_{,\rho\rho}-(B\rho)_{,zz}\right]
         +\frac{1}{4}\,B^3\rho^3e^{-4\nu}\left[(\omega_{,\rho})^2-(\omega_{,z})^2\right],
         \label{eq:EFE4} \\
  & \zeta_{,\rho}(B\rho)_{,z}+\zeta_{,z}(B\rho)_{,\rho}
       = -2B\rho \nu_{,\rho}\nu_{,z}
         -(B\rho)_{,\rho z}+\frac{1}{2}\,B^3\rho^3e^{-4\nu}\omega_{,\rho}\omega_{,z} \,, \label{eq:EFE5}
\end{align}
where $\nabla$ and $\nabla\cdot$ denote gradient and divergence in an (auxiliary) Euclidean three-space.
The last two equations (for $\zeta$) are integrable, provided the first three {\em vacuum} equations hold. The axis boundary condition $e^\zeta\!=\!B$ implies that the $\zeta$ function can elsewhere be obtained according to
$\zeta(r,\theta)=\int_0^\theta\zeta_{,\theta}{\rm d}\zeta+\ln B$,
where $\zeta_{,\theta}$ follows from the last two field equations.

The treatment of the stationary axisymmetric problem (\ref{eq:EFE1})--(\ref{eq:EFE5}) usually starts from a suitable solution of the first equation (\ref{eq:EFE1}). In the parametrization (coordinates) we use here, it is convenient to choose\footnote
{When working in the Weyl-type coordinates ($t$,$\rho$,$z$,$\phi$) and with the corresponding, Weyl-Lewis-Papapetrou form of the metric, the $B$-equation is usually being satisfied by $B\!=\!1$. This choice is advantageous in most respects, but it makes the horizon ``degenerate'' to a central segment of the symmetry axis $\rho\!=\!0$, which is not suitable for discussion of its properties. Different solutions for $B$ are also possible for the ``Carter-Thorne-Bardeen'' form of the metric we use here; however, changing $B$ generally does not imply a real physical difference, it effectively corresponds to a certain re-definition of coordinates, cf. section \ref{grav-potential} (only the $B\!=\!1/\rho$ choice leads to different, plane-wave solutions).}
\begin{equation}
  B = 1-\frac{k^2}{4r^2}\;.
\end{equation}
With such a choice, the horizon lies where $B\!=\!0$, hence on $r\!=\!k/2$. This reveals the meaning of the constant $k$ (which is supposed to be positive), in particular, for Schwarzschild one has $k\!=\!M$; for Kerr it would be $k\!=\!M\!+\!\sqrt{M^2\!-a^2}\,$, with $a$ the centre's specific angular momentum ($k\!=\!0$ would correspond to an extreme black hole, or to a Minkowski space-time).

The main task is to solve the coupled equations (\ref{eq:EFE2}) and (\ref{eq:EFE3}), and then to integrate equations (\ref{eq:EFE4}) and (\ref{eq:EFE5}) for $\zeta$. With the choice $B\!=\!1\!-\!\frac{k^2}{4r^2}$ (thus with $B_{,\theta}\!=\!0$) and written out explicitly in the ($t$,$\rho$,$z$,$\phi$) coordinates, these equations read
\begin{align}
  & (r^2\nu_{,r})_{,r}+r^2\nu_{,r}(\ln B)_{,r}+\nu_{,\theta\theta}+\nu_{,\theta}\cot\theta
       = \frac{B^2 r^2}{2e^{4\nu}}\sin^2\theta\left[r^2(\omega_{,r})^2 + (\omega_{,\theta})^2\right]
         +4\pi r^2 (\sigma + P)\,\frac{1+v^2}{1-v^2}\,\delta(z) \,, \\
  & r^2\omega_{,rr}+4r\omega_{,r}(1-r\nu_{,r})+3r^2\omega_{,r}(\ln B)_{,r}+\omega_{,\theta\theta}
    +3\omega_{,\theta}\cot\theta-4\omega_{,\theta}\nu_{,\theta}
       = -\frac{16\pi re^{2\nu}}{B\sin\theta}(\sigma+P)\,\frac{v}{1-v^2}\,\delta(z) \,, \\
  & (2-B)r\zeta_{,r}-B\zeta_{,\theta}\cot\theta
       = B\left[r^2(\nu_{,r})^2-(\nu_{,\theta})^2\right]+2B-2
         -\frac{1}{4}\,B^3 r^2 e^{-4\nu}\sin^2\theta\left[r^2(\omega_{,r})^2-(\omega_{,\theta})^2\right], \\
  & Br\zeta_{,r}\cot\theta+(2-B)\zeta_{,\theta}
       = 2Br \nu_{,r}\nu_{,\theta}+2(1-B)\cot\theta
         -\frac{1}{2}\,B^3 r^3 e^{-4\nu}\omega_{,r}\omega_{,\theta}\sin^2\theta \;. 
\end{align}

\subsection{Counter-rotating interpretation of thin discs}

When asking about counterbalance to its (and black hole’s) gravity, the disc may either be considered as a solid structure (a set of circular hoops), or, to the contrary, as a non-coherent mix of azimuthal streams. In the astrophysical context, one usually adheres to the latter extreme possibility: that the disc is composed of two non-interacting streams of particles which follow stationary circular orbits in opposite azimuthal directions \citep{bib:MorganMorgan1969,bib:LyndenBell1978,bib:LambHam1989,bib:Bic1993,bib:BicLed1993,bib:KleinRich1999,bib:GonzEspi2003,bib:Garcia2004}. These orbits are geodesic if there is no radial stress acting within the disc ($T^\rho_\rho\!=\!0$). The surface energy-momentum tensor is thus decomposed as
\begin{equation}
   S^{\alpha\beta} = \sigma_+ u_+^\alpha u_+^\beta + \sigma_- u_-^\alpha u_-^\beta \,,
\end{equation}
where the $+/-$ signs indicate the stream orbiting in a positive/negative sense of $\phi$, as taken with respect to the ``average'' fluid four-velocity $u^\alpha$. The four-velocities are of the $u_\pm^\alpha\!=\!u_\pm^\alpha (1,0,0,\Omega_\pm)$ form again, so
\begin{equation}
  S^t_\phi = B\rho e^{-2\nu}\!
             \left(\frac{\sigma_+ v_+}{1-v_+^2}+\frac{\sigma_- v_-}{1-v_-^2}\right), \qquad
  S^\phi_\phi-\omega S^t_\phi = \frac{\sigma_+ v_+^2}{1-v_+^2}+\frac{\sigma_- v_-^2}{1-v_-^2} \;, \qquad
  -S^t_t-\omega S^t_\phi = \frac{\sigma_+}{1-v_+^2}+\frac{\sigma_-}{1-v_-^2} \;,
\end{equation}
now with $\Omega_\pm$ (and the corresponding $v_\pm$) given by free circular motion, i.e. by the roots of equation
$g_{tt,\alpha}+2g_{t\phi,\alpha}\Omega+g_{\phi\phi,\alpha}\Omega^2=0$.
Such a motion is only possible in the equatorial plane ($z\!=\!0$) in general, where just the radial component of acceleration remains relevant, vanishing if
\begin{equation}  \label{Omega_pm}
  \Omega=\Omega_\pm
  =\omega+\frac{g_{\phi\phi}\omega_{,\rho}}{g_{\phi\phi,\rho}}
   \pm\sqrt{\left(\omega+\frac{g_{\phi\phi}\omega_{,\rho}}{g_{\phi\phi,\rho}}\right)^{\!2}
            -\frac{g_{tt,\rho}}{g_{\phi\phi,\rho}}} \;;
\end{equation}
in particular, with the $B\!=\!1$ choice, this expression reduces to
\begin{equation}
  \Omega_\pm
  =\omega+
   \frac{\rho^2\omega_{,\rho}\pm\sqrt{(\rho^2\omega_{,\rho})^2+4e^{4\nu}\rho\nu_{,\rho}(1-\rho\nu_{,\rho})}}
        {2\rho(1-\rho\nu_{,\rho})} \;.
\end{equation}
The interpretation is only possible where the expression under the square root is non-negative; physically, this is not satisfied for discs with ``too much matter on larger radii'': in such a case, the total gravitational pull at a given location points {\em outwards}, so ``no angular velocity is low enough'' to admit Keplerian orbiting there. Parameters of the counter-rotating picture ($\sigma_\pm$,$\Omega_\pm$) and the ``total'', one-stream parameters ($\sigma$,$P$,$\Omega$) are related by comparing the respective two forms of the energy-momentum tensor,
$\sigma u^\alpha u^\beta+P w^\alpha w^\beta = \sigma_+ u_+^\alpha u_+^\beta+\sigma_- u_-^\alpha u_-^\beta\,$.
From the trace of this equation and from its projections onto $u_\alpha u_\beta$ and $w_\alpha w_\beta$, while using a suitable expression of the scalar products
\begin{equation}
  u_\pm^\alpha u_\alpha = \frac{v v_\pm-1}{\sqrt{1-v^2}\,\sqrt{1-v_\pm^2}} \;, \qquad
  u_\pm^\alpha w_\alpha = \frac{v_\pm - v}{\sqrt{1-v^2}\,\sqrt{1-v_\pm^2}} \;, \qquad
  g_{\alpha\beta} u_+^\alpha u_-^\beta = \frac{v_+ v_- - 1}{\sqrt{1-v_+^2}\,\sqrt{1-v_-^2}} \;,
\end{equation}
one finds
\begin{equation}
  \sigma-P = \sigma_+ + \sigma_- \,, \qquad
  \sigma = \frac{\sigma_+\,(vv_+-1)^2}{(1-v^2)(1-v_+^2)}+
           \frac{\sigma_-\,(vv_--1)^2}{(1-v^2)(1-v_-^2)} \;, \quad
  P = \frac{\sigma_+\,(v_+-v)^2}{(1-v^2)(1-v_+^2)}+
      \frac{\sigma_-\,(v_--v)^2}{(1-v^2)(1-v_-^2)} \;,
\end{equation}
\begin{equation}  \label{P/sigma}
  \frac{P}{\sigma}
    = \frac{u_+^\alpha w_\alpha}{u_+^\beta u_\beta}\,
      \frac{u_-^\lambda w_\lambda}{u_-^\kappa u_\kappa}
    = \frac{(v_+-v)(v-v_-)}{(1-vv_+)(1-vv_-)}
  \qquad \Longrightarrow \qquad
  \frac{\sigma+P}{\sigma_+ +\sigma_-} = \frac{1+\frac{P}{\sigma}}{1-\frac{P}{\sigma}} = \dots \,,
\end{equation}
\begin{equation}
  \sigma_\pm (u^t_\pm)^2 = \pm\frac{S^{t\phi}-\Omega_\pm S^{tt}}{\Omega_+ -\Omega_-}
  \qquad \Longrightarrow \qquad
  \frac{\sigma_\pm}{1-v_\pm^2} = \pm\frac{\sigma(v_\pm-v)-Pv(1-vv_\pm)}{(1-v^2)(v_+ -v_-)} \;.
\end{equation}
We may also, for example, compare expressions for $S_{\alpha\beta}S^{\alpha\beta}$, finding the relation
\begin{equation}
  \sigma^2+P^2 = \sigma_+^2+\sigma_-^2+2\sigma_+\sigma_-(g_{\alpha\beta}u_+^\alpha u_-^\beta)^2
  =\sigma_+^2+\sigma_-^2+ 2\sigma_+\sigma_-\,\frac{(1-v_+ v_-)^2}{(1-v_+^2)(1-v_-^2)} \;,
\end{equation}
which in combination with $\sigma\!-\!P=\sigma_+\!+\sigma_-$ also leads to
\begin{equation}
  \sigma P = \sigma_+\sigma_-\,\frac{(v_+-v_-)^2}{(1-v_+^2)(1-v_-^2)} \;,
  \qquad
  (\sigma_+ -\sigma_-)^2 = (\sigma+P)^2-4\sigma P\,\frac{(1-v_+v_-)^2}{(v_+-v_-)^2} \;.
\end{equation}

\subsection{One-stream and two-stream interpretations: integrating jumps in the field equations}
\label{jumps}

Relation between the jumps of $g_{\alpha\beta,\mu}$ across the disc and $S^\alpha_\beta$ are obtained by integrating the field equations (\ref{eq:EFE1})--(\ref{eq:EFE+}) over the infinitesimal interval $\langle z\!=\!0^-,z\!=\!0^+ \rangle$. Only the terms proportional to $\delta(z)$ (i.e. the source terms on the right-hand sides and the terms linear in $B_{,zz}$, $\nu_{,zz}$, $\omega_{,zz}$ and $\zeta_{,zz}$ on the left-hand side) contribute according to
$\int_{z=0^-}^{z=0^+}\!\nu_{,zz}{\rm d}z = 2\nu_{,z}(z\!=\!0^+)$ (etc.), so we have $B_{,z}(z\!=\!0^+)=0$ and
\begin{align}
  \nu_{,z}(z\!=\!0^+)
    &= 2\pi\left(S^\phi_\phi-2\omega S^t_\phi-S^t_t\right)
     = 2\pi (\sigma+P)\,\frac{1+v^2}{1-v^2}
     = 2\pi\left(\sigma_+\,\frac{1+v_+^2}{1-v_+^2}+\sigma_-\,\frac{1+v_-^2}{1-v_-^2}\right),
     \label{nuz,jump} \\
  \omega_{,z}(z\!=\!0^+)
    &=-\frac{8\pi\,S^t_\phi}{B^2\rho^2 e^{-4\nu}}
     = -8\pi (\sigma+P)\,\frac{\Omega-\omega}{1-v^2}
     = -8\pi\left(\sigma_+\,\frac{\Omega_+ -\omega}{1-v_+^2}+
                  \sigma_-\,\frac{\Omega_- -\omega}{1-v_-^2}\right),
     \label{omegaz,jump} \\
  \zeta_{,z}(z\!=\!0^+)
    &= 4\pi\left(S^\phi_\phi-\omega S^t_\phi\right)
     = 4\pi\,\frac{\sigma v^2+P}{1-v^2}
     = 4\pi\left(\frac{\sigma_+ v_+^2}{1-v_+^2}+\frac{\sigma_- v_-^2}{1-v_-^2}\right),
     \label{zetaz,jump}
\end{align}
where we have expressed the results in terms of the one-stream as well as two-stream form of $S^\alpha_\beta$.

\section{Perturbation scheme}
\label{ch:Perturbation}

We will look for a solution of equations (\ref{eq:EFE2}) and (\ref{eq:EFE3}) in the form of series, expanding
\begin{equation} \label{eq:Expansion}
  \nu = \sum\limits_{j=0}^\infty \nu_j \lambda^j \,, \qquad 
  \omega = \sum\limits_{j=0}^\infty \omega_j \lambda^j \,, \qquad
  \zeta = \sum\limits_{j=0}^\infty \zeta_j \lambda^j \,,
\end{equation}
where the coefficients $\nu_j$, $\omega_j$ and $\zeta_j$ depend on $r$ and $\theta$ (or $\rho$ and $z$) and the dimensionless parameter $\lambda$ is proportional to the ratio of the disc mass to the black-hole mass $M$. More specifically, let it be related by 
\begin{equation}  \label{eq:Density}
  (\sigma+P)\delta(z)
  \equiv \lambda\Sigma(\rho)\delta(z)
  =\lambda\Sigma(r)\,\frac{1}{r}\,\delta(\cos\theta)
  =-\lambda\Sigma(r)\,\frac{1}{r}\,\delta(\theta\!-\!\pi/2),
\end{equation}
where $\delta$ denotes the $\delta$-distribution and $\lambda\Sigma\!\equiv\!\sigma\!+\!P$ is an ``effective'' surface density. The functions $\nu_0$, $\omega_0$ and $\zeta_0$ represent the black-hole background, i.e. the Schwarzschild metric which in isotropic coordinates (recall that $\rho\!=\!r\sin\theta$ and $z\!=\!r\cos\theta$) reads
\begin{equation}
  {\rm d}s^2
  = -\left(\frac{2r-M}{2r+M}\right)^2 {\rm d}t^2
    +\left(1+\frac{M}{2r}\right)^4 \left({\rm d}r^2+r^2{\rm d}\theta^2+r^2\sin^2\theta\,{\rm d}\phi^2\right), 
\end{equation}
hence
\begin{equation}
  \nu_0 = \ln\frac{2r-M}{2r+M} \,, \qquad
  \omega_0 = 0, \qquad
  B = e^{\zeta_0} = 1-\frac{M^2}{4r^2} \;,
\end{equation}
and the corresponding orbital velocity is
\begin{equation}
  \Omega_0 =  \frac{8\,\sqrt{Mr^3}}{(2r+M)^3} \,, \qquad
  v_0 = \frac{2\,\sqrt{Mr}}{2r-M} \,.
\end{equation}

Substituting (\ref{eq:Expansion}) and (\ref{eq:Density}) into (\ref{eq:EFE2}) and (\ref{eq:EFE3}) and subtracting the pure-Schwarzschild terms, one obtains
\begin{align}  \label{eq:PertEFE2}
  \sum\limits_{k=1}^\infty \lambda^k \nabla\!\cdot\!(B\nabla\nu_k)
  = & \sum\limits_{k=2}^\infty \lambda^k
      \left\{\frac{(2r+M)^7\sin^2\theta}{2^7 r^4(2r-M)}\;
             \sum\limits_{l=0}^{k-2}
             \left[{\rm exp}\bigg(\!{-}4\sum\limits_{j=1}^\infty \lambda^j \nu_j\!\bigg)\right]_l
             \sum\limits_{m=1}^{k-l-1}\!\nabla\omega_m\cdot\nabla\omega_{k-l-m}\right\} + \nonumber \\
    &+4\pi\frac{B}{r}\,\Sigma\,\delta(\cos\theta)
      \;\sum\limits_{k=0}^\infty \lambda^{k+1} \left[\frac{1+v^2}{1-v^2}\right]_k,
\end{align}
\begin{align}  \label{eq:PertEFE3}
  \sum\limits_{k=1}^\infty \lambda^k
  \nabla\!\cdot\!\left\{\frac{(2r+M)^7\sin^2\theta}{2^6 r^4(2r-M)}\;\nabla\omega_k\right\}
  = &-\sum\limits_{k=2}^\infty \lambda^k
      \sum\limits_{l=1}^{k-1}
      \nabla\!\cdot\!
      \left\{\frac{(2r+M)^7\sin^2\theta}{2^6 r^4(2r-M)}
             \left[{\rm exp}\bigg(\!{-}4\sum\limits_{j=1}^\infty \lambda^j \nu_j\!\bigg)\right]_{k-l}
             \!\!\!\!\nabla\omega_l\right\} - \nonumber \\
    &-16\pi\left(1+\frac{M}{2r}\right)^4\Sigma\,\delta(\cos\theta)
      \;\sum\limits_{k=0}^\infty \lambda^{k+1} \left[\frac{v}{1-v^2}\right]_k,
\end{align}
where $[f]_k$ means the coefficient standing at $\lambda^k$ in Taylor expansion of $f$.
Since the background is static, the first-order equations only contain mass-energy terms multiplied by the background metric on the right-hand sides, because the first-line sums do not contribute. In higher orders, the right-hand sides contain only lower-order terms. (For non-static backgrounds the equations do not decouple so easily.)

Now, the eigen-functions with respect to $\theta$ of the operator on the left-hand sides of (\ref{eq:PertEFE2}) and (\ref{eq:PertEFE3}) are the Legendre polynomials $P_l(\cos\theta)$ and the Gegenbauer polynomials $C^{(3/2)}_l(\cos\theta)$, respectively.\footnote
{In Will's article the latter are denoted by $T^{3/2}_l(\cos\theta)$.}
Introducing a dimensionless radius $x\!:=\!\frac{r}{M}\!\left(1\!+\!\frac{M^2}{4r^2}\right)$, we may thus write 
\begin{equation}  \label{eq:HarmonicExpansion}
  \nu_l = \sum\limits_{j=0}^\infty \nu_{lj}(x)\,P_j(\cos\theta), \qquad
  \omega_l = \sum\limits_{j=0}^\infty \omega_{lj}(x)\,C^{(3/2)}_j(\cos\theta) \,.
\end{equation}
Substituting this into (\ref{eq:PertEFE2}) and (\ref{eq:PertEFE3}) leads to 
\begin{align}
  \sum\limits_{j=0}^\infty
  \left\{\frac{{\rm d}}{{\rm d}x}\left[(x^2-1)\,\frac{{\rm d}\nu_{lj}}{{\rm d}x}\right]-j(j+1)\nu_{lj}\right\}
  P_j(\cos\theta)
  & = R_l(x,\theta) \,, \label{eq:PertEFE2_1} \\
  \sum\limits_{j=0}^\infty
  \left\{(x^2-1)\,\frac{{\rm d}}{{\rm d}x}\left[(x+1)^4\,\frac{{\rm d}\omega_{lj}}{{\rm d}x}\right]
         -(x+1)^4 j(j+3)\omega_{lj}\right\}
  C^{(3/2)}_j(\cos\theta)
  & = S_l(x,\theta) \,, \label{eq:PertEFE3_1}
\end{align}
where $R_l(x,\theta)$ and $S_l(x,\theta)$ stand, up to an $l$-independent multiplication factor, for the coefficients of expansion (with respect to $\lambda$) of the right-hand sides of equations (\ref{eq:PertEFE2}) and (\ref{eq:PertEFE3}); specifically,
\begin{equation}
  \sum\limits_{l=0}^\infty R_l(x,\theta)\lambda^l
  = \frac{r^2}{B} \left[{\rm r.h.s. \; of \; (\ref{eq:PertEFE2})}\right], \qquad
  \sum\limits_{l=0}^\infty S_l(x,\theta)\lambda^l
  = \frac{Br^4}{M^4\sin^2\theta} \left[{\rm r.h.s. \; of \; (\ref{eq:PertEFE3})}\right].
\end{equation}
Provided that $R_l$ and $S_l$ do not diverge on the axis $\theta\!=\!0,\,\pi$, these coefficients can be also decomposed as
\begin{equation}
  R_l(x,\theta) = \sum\limits_{j=0}^\infty R_{lj}(x)\,P_j(\cos\theta) \,, \qquad
  S_l(x,\theta) = \sum\limits_{j=0}^\infty S_{lj}(x)\,C^{(3/2)}_j(\cos\theta) \,,
\end{equation}
where 
\begin{align}
  R_{lj}(x) &= \frac{2j+1}{2}
               \int\limits_{-1}^1 R_l(x,\theta)\,P_j(\cos\theta)\;{\rm d}(\cos\theta) \,, \\
  S_{lj}(x) &= \frac{2j+3}{2(j+1)(j+2)}
               \int\limits_{-1}^1 S_{l}(x,\theta)\,C^{(3/2)}_j(\cos\theta)\sin^2\theta\;{\rm d}(\cos\theta) \;.
\end{align}

Demanding that the equations (\ref{eq:PertEFE2_1}) and (\ref{eq:PertEFE3_1}) hold for each order (multipole moment) separately, one obtains a system of independent ordinary differential equations
\begin{align}
  \frac{{\rm d}}{{\rm d}x}\left[(x^2-1)\,\frac{{\rm d}\nu_{lj}}{{\rm d}x}\right]
     -j(j+1)\,\nu_{lj}
  & = R_{lj} \,, \label{eq:EFE2_Dec} \\
  (x^2-1)\,\frac{{\rm d}}{{\rm d}x}\left[(x+1)^4\,\frac{{\rm d}\omega_{lj}}{{\rm d}x}\right]
     -(x+1)^4 j(j+3)\,\omega_{lj}
  & = S_{lj} \,, \label{eq:EFE3_Dec}
\end{align}
where $l\in\mathbb{N}$, $j\in\mathbb{Z}^+_0$.
They only contain lower-order source terms (assumed to be given) for every $l$ and are to be supplemented by boundary conditions on the horizon $x\!=\!1$ ($\nu_{lj}$ and $\omega_{lj}$ are supposed to be regular there) and at spatial infinity ($\nu_{lj}$ and $\omega_{lj}$ should vanish there).
Of many techniques available for such equations, we shall focus on finding their Green functions.

Let us start from fundamental systems of equations (\ref{eq:EFE2_Dec}) (\ref{eq:EFE3_Dec}). The first one is the Legendre differential equation whose fundamental system can be expressed as linear combination of Legendre functions of the first and second kinds
\begin{equation}  \label{eq:QDef}
  P_j(x) \qquad {\rm and} \qquad
  Q_j(x) = P_j(x) \int\limits_x^\infty \frac{{\rm d}\xi}{(\xi^2-1)\left[P_j(\xi)\right]^2} \;.
\end{equation}
For the second equation, \cite{bib:Will1974} used the substitution $t\!=\!(x+1)/2$ which transforms it to the hypergeometric differential equation, having two generators of the fundamental system,
\begin{equation}  \label{eq:GDef}
  F_j(x) = {}_2F_1\!\left(-j,j+3;4;\frac{x+1}{2}\right) \qquad {\rm and} \qquad
  G_j(x) = F_j(x) \int\limits_x^\infty \frac{{\rm d}\xi}{(\xi+1)^4 \left[F_j(\xi)\right]^2} \;, 
\end{equation} 
where ${}_2F_1(a,b;c;\xi)$ denotes the Gauss hypergeometric function. Note that since $j\in\mathbb{Z}^+_0$, $F(x)$ is in fact a polynomial of degree $j$. 
Asymptotically (as $x\!\rightarrow\!\infty$), $P_j(x)\!\sim\!x^j$, $Q_j(x)\!\sim\!x^{-l-1}$, $F(x)\sim\!x^j$ and $G(x)\!\sim x^{-j-3}$. At the horizon ($x\!=\!1$), $Q_j(x)$ and $G_j(x)$ diverge (except $G_0(x)$ which will be discussed later).

Given the above boundary conditions, the Green functions of equations (\ref{eq:PertEFE2_1}), (\ref{eq:PertEFE3_1}) can be found in the form
\begin{align}
  {\cal G}^\nu_j(x,x') &=
      \left\{\begin{matrix}
             \;-Q_j(x)\,P_j(x') & \;\;{\rm for}\;\; & x\geq x' \cr
             \;-P_j(x)\,Q_j(x') & \;\;{\rm for}\;\; & x\leq x'
             \end{matrix}\right. \,, \label{eq:GFPertDecNu}\\
  {\cal G}^\omega_j(x,x') &=
      \left\{\begin{matrix}
             \;-G_j(x)\,F_j(x') & \;\;{\rm for}\;\; & x\geq x' \cr
             \;-F_j(x)\,G_j(x') & \;\;{\rm for}\;\; & x\leq x'
             \end{matrix}\right. \,, \label{eq:GFPertDecOmega}
\end{align}
and their inhomogeneous solutions as
\begin{align}
  \nu_{lj}(x) &= \int\limits_1^\infty R_{lj}(x')\,{\cal G}^\nu_j(x,x')\;{\rm d}x' \,,
  \label{eq:EFE2_DecInhomogSol} \\
  \omega_{lj}(x) &= \int\limits_1^\infty S_{lj}(x')\,{\cal G}^\omega_j(x,x')\;{\rm d}x'
                    +\frac{J_l\,\delta_j^0}{(x+1)^3} \;, \label{eq:EFE3_DecInhomogSol}
\end{align}
where $J_l$ are arbitrary constants, representing a choice of the black-hole spin (see section \ref{sec:WillComp}).
Such a solution is unique up to a coordinate transformation. Note that one could add to the Green functions terms proportional to $P_0(x)\!=\!F_0(x)\!=\!1$ which is everywhere finite, but such an addition only corresponds to a rescaling of time and (thus) of the coordinate angular velocity, so it has no invariant physical effect.

Using (\ref{eq:HarmonicExpansion}), we may write
\begin{align}
  \nu_l(x,\theta) 
    &= \sum\limits_{j=0}^\infty \nu_{lj}(x)\,P_j(\cos\theta)
     = \int\limits_1^\infty
       \sum\limits_{j=0}^\infty R_{lj}(x')\,{\cal G}^\nu_j(x,x')\,P_j(\cos\theta)
       \;{\rm d}x' 
     = \int\limits_1^\infty \int\limits_{-1}^1
       R_l(x',\theta')\,{\cal G}^\nu(x,\theta,x',\theta')\;{\rm d}x'\,{\rm d}(\cos\theta') \,, 
    \label{eq:GFNuFull} \\
  \omega_l(x,\theta)
    &= \sum\limits_{j=0}^\infty \omega_{lj}(x)\,C^{(3/2)}_j(\cos\theta)
     = \int\limits_1^\infty
       \sum\limits_{j=0}^\infty S_{lj}(x')\,{\cal G}^\omega_j(x,x')\,C^{(3/2)}_j(\cos\theta)
       \;{\rm d}x'
     = \int\limits_1^\infty \int\limits_{-1}^1
       S_l(x',\theta')\,{\cal G}^\omega(x,\theta,x',\theta')\;{\rm d}x'\,{\rm d}(\cos\theta') \,,
    \label{eq:GFOmegaFull}
\end{align}
where
\begin{align}
  {\cal G}^\nu(x,\theta,x',\theta')
  &:= -\sum\limits_{j=0}^\infty \frac{2j+1}{2}\,
       P_j(\min(x,x'))\,Q_j(\max(x,x'))\,P_j(\cos\theta)\,P_j(\cos\theta')) \,,
       \label{eq:GFNuDef} \\
  {\cal G}^\omega(x,\theta,x',\theta')
  &:= -\sum\limits_{j=0}^\infty \frac{2j+3}{2(j+1)(j+2)}\,
       F_j(\min(x,x'))\,G_j(\max(x,x'))\,C^{(3/2)}_j(\cos\theta)\,C^{(3/2)}_j(\cos\theta')
       \quad \left[+\frac{J_l}{(x+1)^3}\right]
       \label{eq:GFOmegaDef}
\end{align}
are Green's functions of homogeneous parts of equations (\ref{eq:PertEFE2}) and (\ref{eq:PertEFE3}). These represent a perturbation by an infinitesimal (2D) circular ring placed at $x'$, $\theta'$ in the first order in $\lambda$. (We omit the $J_l$ term in the following, see below.)

\subsection{Differences from the Will's article}
\label{sec:WillComp}

\cite{bib:Will1974} found the Green functions of equations (\ref{eq:GFPertDecNu}) and (\ref{eq:GFPertDecOmega}) or, more precisely, he proposed the inhomogeneous solutions (\ref{eq:EFE2_DecInhomogSol}) and (\ref{eq:EFE3_DecInhomogSol}). He employed {\em three} expansions of the solutions: with respect to the linear mass density of the ring, with respect to the angular velocity of the horizon, and the multipole expansion performed to convert partial differential equations into the ordinary ones. In contrast, we use only {\em two} expansions: with respect to the disc density and the multipole expansion. Below (see section \ref{ch:GF}), we will even be able to drop the multipole expansion and work solely with expansion with respect to the disc (ring) density. Namely, the rotational expansion can be ``reconstructed'' using the constants $J_l$ (not employed in the Will's paper). This is possible due to $F_j(1)\!=\!0$ for $j\!>\!0$ (see, for example, (\ref{eq:CDDifForm})), which implies that higher multipole moments do not contribute to the black-hole angular velocity at all, the only important entering terms (in a given order $\lambda^l$ of the mass perturbation) being proportional to $F_0(x)\!=\!1$, $G_0(x)\!=\!(x+1)^{-3}$ or $\omega_{l0}(x)$, where $\omega_{l0}$ is the inhomogeneous solution of (\ref{eq:EFE3_Dec}) given by (\ref{eq:EFE3_DecInhomogSol}) (with $J_l\!=\!0$). The first term $F_0(x)\!=\!1$ only adds constant coordinate angular velocity, so it is not a physical degree of freedom (it can be cancelled out by a coordinate transformation). The meaning of the remaining two terms is revealed by their behaviour in a vicinity of the disc (ring): $G_0$ is smooth everywhere, but $\omega_{l0}$ jumps in its first derivative (in the disc case it thus contributes to the density of the disc energy-momentum). Hence, the term proportional to $G_0$ can be interpreted as the black hole's ``own'' angular velocity (perturbation towards the Kerr solution), while the terms proportional to $\omega_{l0}$ represent rotational perturbation due to the presence of the rotating disc (ring). Needless to say, such an interpretation is not unique, because $\omega_{l0}+C_l G_0$ ($C_l$ are arbitrary constants) is also a solution of (\ref{eq:EFE3_Dec}) contributing to the source. Since there is no clear way how to say which choice of $C_l$ is the ``correct'' one, we will adhere to $C_l\!=\!0$ for simplicity.

Having prescribed the black-hole angular velocity
\begin{equation}
  \omega_{\rm H} = \sum\limits_{l=1}^\infty \beta_l \lambda^l \,,
\end{equation}
one can find the $J_l$ constants according to
\begin{equation}
  J_l = 8\left[\beta_l-\left.\omega_{l0}(1)\right|_{J_l=0}\right]
      = 8\beta_l + \int\limits_{1}^\infty \frac{8S_{l0}(x')}{(x'+1)^3}\;{\rm d} x' \,,
\end{equation}
where $\omega_{lj}|_{J_l=0}$ stands for the right-hand side of (\ref{eq:EFE3_DecInhomogSol}) with $J_l\!=\!0$, and $S_{l0}$ contains lower-order terms of the expansion in $\lambda$ (i.e., it also contains $J_k$ with $k\!<\!l$).

\section{Convergence and related issues}
\label{ch:Convergence}

The main attribute of the above procedure is its speed of convergence to the desired solution. The problem is familiar from spectral methods (multipole expansion can actually be viewed as a spectral method): the numerics works well when the desired result is smooth, otherwise convergence problems arise. More specifically, one can expect exponential convergence for analytical function, whereas at most a power-law one for functions having some derivatives discontinuous (see, for example, \citealt{bib:GrandNovak2008}).

Let us begin with the potential for the ring case. Focusing on the region {\em outside} the black hole, $1\!\leq\!x_<\!\leq\!x_>$, and regarding the asymptotic behaviour of special functions (see, for example, \citealt{bib:DLMF}), one can, after lengthy calculations, find that
\begin{equation}
  \frac{2j+1}{2}\;P_j(\cos\theta)\,P_j(x_<)\,Q_j(x_>)\,P_j(0)
  \approx \frac{1}{j} \left(\frac{x_< +\sqrt{x_<^2-1}}{x_> +\sqrt{x_>^2-1}}\right)^{\!\frac{1}{2}+j}
          \left[f(x,x')+O(1/j)\right] \,,
\end{equation}
where $f(x,x')$ is a suitable function independent of $j$, $x_<:=\min(x,x')$ and $x_>:=\max(x,x')$. This implies that the sum (\ref{eq:GFNuFull}) exponentially converges outside the ring radius and {\em conditionally} converges (like $1/j$) just on the radius of the ring. Practically, this means that near the ring radius one has to include quite many terms (in $j$) in order to get reasonably small oscillations. (Convergence could be expected to be slow near a singular source.)

In the disc case, the situation is better and worse at the same time. After a lengthy calculation again, one finds that
\begin{equation}  \label{eq:ConvDisc}
  \left|
  \int\limits_{x_{{\rm R}-}}^{x_{{\rm R}+}}
  \frac{2j+1}{2}\,R_l(x_{\rm R})\,P_j(\cos\theta)\,P_j(x_<)\,Q_j(x_>)\,P_j(0)\;{\rm d}x_{\rm R}
  \right|
  <
  \frac{1}{j^2} \left(\frac{x'_< +\sqrt{x'^2_< -1}}{x'_> +\sqrt{x'^2_> -1}}\right)^{\!\frac{1}{2}+j}
  \left[f(x,x')+O(1/j)\right] \,,
\end{equation}
where $x_<$ and $x_>$ have the same meaning as above, i.e. $x_<:=\min(x,x_{\rm R})$ and $x_>:=\max(x,x_{\rm R})$, and $x'_<$, $x'_>$ stand for analogous quantities taken with respect to the nearest radius lying inside the disc ($x'_{\rm R}$), i.e.,
\begin{equation}
  x'_< = \left\{\begin{matrix}
                 x & {\rm when}\; & x \leq x_{\rm out} \cr
                 x_{\rm out} & {\rm when}\; & x \geq x_{\rm out}
                 \end{matrix}\right. ,
  \qquad
  x'_> = \left\{\begin{matrix}
                 x_{\rm in} & {\rm when}\; & x \leq x_{\rm in} \cr
                 x & {\rm when}\; & x \geq x_{\rm in}
             \end{matrix}\right. ,
\end{equation}
where $x_{\rm in}$ and $x_{\rm out}$ represent inner and outer rim of the disc. 

Equation (\ref{eq:ConvDisc}) factually yields a lower bound of the convergence speed. (One can already obtain a speed estimate by checking the simplest example, i.e. constant $S_{lj}$.)
We can thus conclude that in the disc case the convergence is exponentially fast outside the radii of the disc and polynomial (like $j^{-2}$) at the radii ``within'' the disc. Therefore, it is generally impossible to improve the polynomial convergence using interpolation in some infinitesimally small area.

Calculation of dragging ($\omega_{jl}$) shows similar behaviour, i.e. exponential convergence outside the radii of the ring (disc) and polynomial one like $j^{-1}$ ($j^{-2}$) at the radii of the source. It is somewhat longer to prove this, so let us only say that one can proceed as follows:
(i) prove that ${\rm d}D^{(\gamma)}_n(x)/{\rm d}x=-2\gamma\,D^{(\gamma+1)}_{n-1}(x)$, where $D^{(\gamma)}_n(x)$ is the second independent solution of the ultraspherical differential equation (the ``second Gegenbauer function'', see section \ref{ch:GF} below),
(ii) express $F_j(x)$ and $G_j(x)$ using (\ref{eq:CDDifForm}), and
(iii) consider the asymptotic behaviour of the Gegenbauer functions.

There is one more challenge for numerical precision: integrals (\ref{eq:QDef}) and (\ref{eq:GDef}), when expressed using elementary functions, have a form of two terms almost cancelling each other.
This problem can be managed by recalling that the Legendre function of the second kind $Q_j(x)$ can be expressed in terms of elementary functions and then evaluated efficiently. Analogously, $G_j(x)$ can be expressed as 
\begin{equation}  \label{eq:GExpansion}
  G_j(x) = \frac{(-1)^j (j+2)!\,(j+3)!}{48\,(2j+3)!}\left(\frac{2}{x+1}\right)^{j+3}
           {}_2F_1\!\left(j,j+3;2j+4;\frac{2}{x+1}\right)
\end{equation}
which converges very well after the Gauss hypergeometric function is expanded suitably.
Let us add that we do not know how to prove the equivalence of (\ref{eq:GExpansion}) and (\ref{eq:GDef}) {\em directly}. However, it is possible to check that both expressions of $G_j(x)$ solve the homogeneous part of (\ref{eq:EFE3_Dec}) and that both vanish at $x\!\rightarrow\!\infty$. Equation (\ref{eq:EFE3_Dec}) is a second-order ordinary linear differential equation, so its fundamental space is two-dimensional. And since the other solution $F_j(x)$ does not vanish at infinity, both expressions of $G_j(x)$ have to be proportional to each other. Comparing their leading terms (of expansion in $1/x$), one finally concludes that the functions exactly coincide.

\section{Green functions}
\label{ch:GF}

\begin{figure}
\begin{center}
\includegraphics[width=0.9\textwidth]{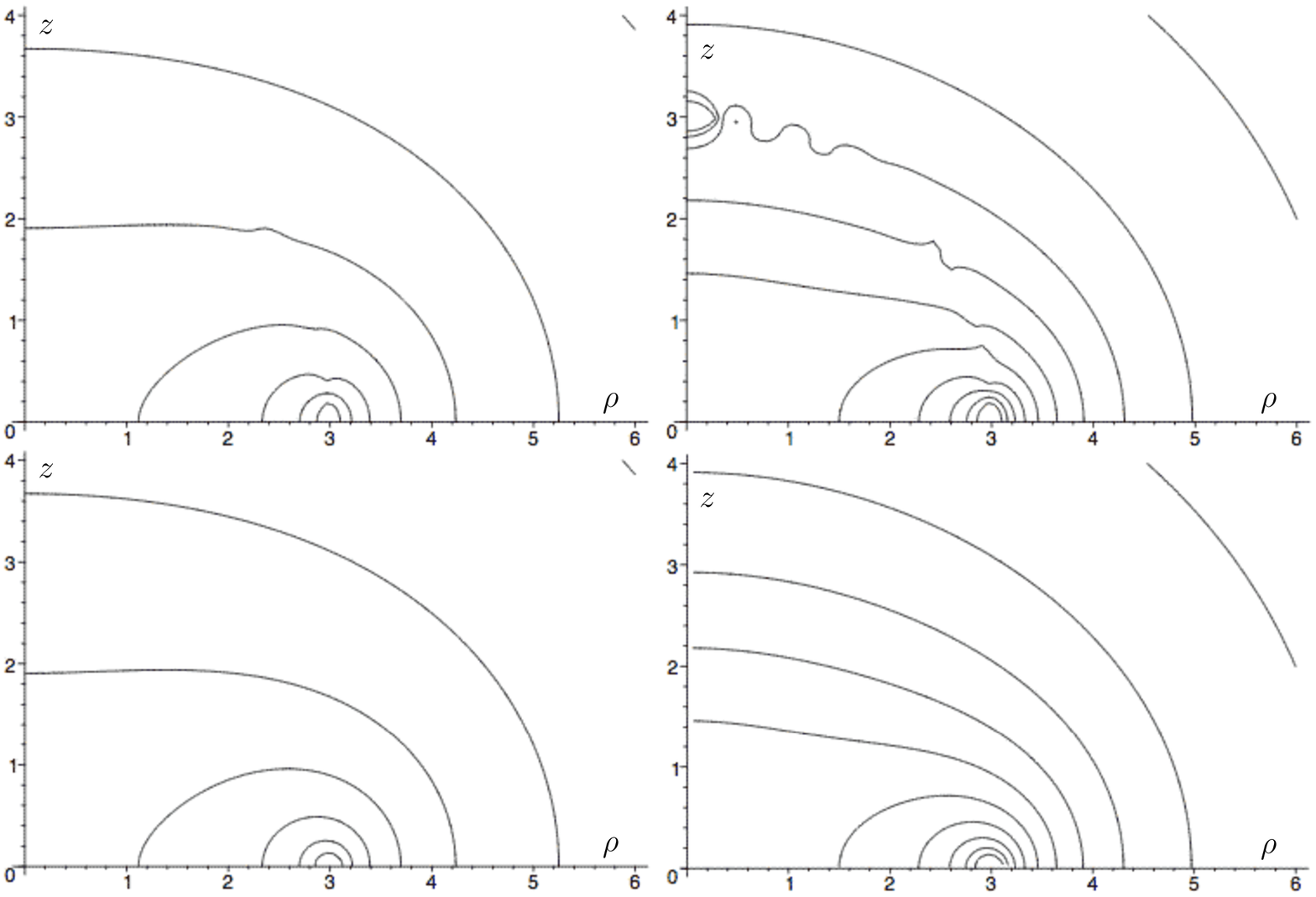}
\end{center}
\caption
{Meridional-plane contours of the gravitational potential ($\nu_1$; left column) and of the linear dragging term ($\omega_1$; right column), generated by a thin ring of matter located at $r\!=\!3{\cal M}$ and having rest mass $0.01{\cal M}$ (where ${\cal M}$ is some mass scale; note that here the disc is the only source, there is no black hole). In the first row, the metric functions are calculated summing the first 30 orders of multipole expansion, yet still there are obvious problems near the ring radius. In the second row, the same are computed (with a clearly better accuracy) using the closed-form Green functions derived in section \ref{ch:GF}. The coordinate axes are given in the units of $M$.}
\label{fig:GF_Exp30}
\end{figure}

To overcome the convergence problems, one can try to avoid angular expansion. A straightforward way to do so is to express the Green functions ${\cal G}^\nu(x,\theta,x',\theta')$ and ${\cal G}^\omega(x,\theta,x',\theta')$ in a closed form. This section is devoted to this task and uses relations which we will justify in detail elsewhere \citep{bib:math-part}.
As an advertisement for the usage of the closed-form Green functions, see figure \ref{fig:GF_Exp30} where, on an example of the ring perturbation, results are compared computed i) by the usual expansions and ii) using the closed-form Green functions. The first 30 terms of multipole expansion are summed there, but the closed-form Green functions provide better results. See also Tables \ref{tab:GFNuConv} and \ref{tab:GFOmegaConv} where the convergence of the multipole expansion towards exact result (given by the closed-form Green functions) is illustrated. One should admit that the multipole series of this type can usually be re-summed in such a way that their convergence is much better even close to the source radius. Anyway, the main advantage of the closed-form Green functions is that one can better integrate them over the source, as exemplified in section \ref{ch:Disc} below.

\begin{table}
\centering
\begin{tabular}{|c|c|c c c c c|c|}
\hline
\multirow{2}{*}{$\theta$} & \multirow{2}{*}{$x$} & \multicolumn{5}{|c|}{number of terms added in (\ref{eq:GFNuDef})} & \multirow{2}{*}{exact value}\\
\cline{3-7} & & 10 & 20 & 30 & 40 & 50 & \\
\hline
\multirow{7}{*}{$\frac{\pi}{2}$} 
& $2.00$ & $-0.19807$ & $-0.19813$ & $-0.19813$ & $-0.19813$ & $-0.19813$ & $-0.19813$ \\
& $2.50$ & $-0.22468$ & $-0.22664$ & $-0.22680$ & $-0.22682$ & $-0.22682$ & $-0.22682$ \\
& $2.90$ & $-0.26750$ & $-0.28772$ & $-0.29629$ & $-0.30058$ & $-0.30293$ & $-0.30636$ \\
& $3.00$ & $-0.28370$ & $-0.31884$ & $-0.34032$ & $-0.35583$ & $-0.36797$ & $-\infty$  \\
& $3.10$ & $-0.25857$ & $-0.27846$ & $-0.28700$ & $-0.29134$ & $-0.29374$ & $-0.29738$ \\
& $3.50$ & $-0.19552$ & $-0.19818$ & $-0.19849$ & $-0.19853$ & $-0.19854$ & $-0.19854$ \\
& $4.00$ & $-0.15445$ & $-0.15478$ & $-0.15479$ & $-0.15479$ & $-0.15479$ & $-0.15479$ \\
\hline
\multirow{7}{*}{$\frac{\pi}{3}$} 
& $2.00$ & $-0.17340$ & $-0.17344$ & $-0.17344$ & $-0.17344$ & $-0.17344$ & $-0.17344$ \\
& $2.50$ & $-0.16688$ & $-0.16747$ & $-0.16754$ & $-0.16755$ & $-0.16755$ & $-0.16755$ \\
& $2.90$ & $-0.15563$ & $-0.15601$ & $-0.15780$ & $-0.15881$ & $-0.15894$ & $-0.15858$ \\
& $3.00$ & $-0.15196$ & $-0.15038$ & $-0.15329$ & $-0.15671$ & $-0.15813$ & $-0.15585$ \\
& $3.10$ & $-0.15012$ & $-0.15044$ & $-0.15221$ & $-0.15323$ & $-0.15337$ & $-0.15299$ \\
& $3.50$ & $-0.14007$ & $-0.14071$ & $-0.14085$ & $-0.14086$ & $-0.14086$ & $-0.14086$ \\
& $4.00$ & $-0.12584$ & $-0.12599$ & $-0.12600$ & $-0.12600$ & $-0.12600$ & $-0.12600$ \\
\hline
\multirow{7}{*}{$\frac{\pi}{4}$} 
& $2.00$ & $-0.16020$ & $-0.16016$ & $-0.16016$ & $-0.16016$ & $-0.16016$ & $-0.16016$ \\
& $2.50$ & $-0.15061$ & $-0.15003$ & $-0.15008$ & $-0.15008$ & $-0.15008$ & $-0.15008$ \\
& $2.90$ & $-0.14321$ & $-0.13879$ & $-0.14019$ & $-0.14108$ & $-0.14074$ & $-0.14062$ \\
& $3.00$ & $-0.14178$ & $-0.13419$ & $-0.13694$ & $-0.14025$ & $-0.13889$ & $-0.13816$ \\
& $3.10$ & $-0.13822$ & $-0.13388$ & $-0.13526$ & $-0.13617$ & $-0.13582$ & $-0.13570$ \\
& $3.50$ & $-0.12655$ & $-0.12582$ & $-0.12591$ & $-0.12592$ & $-0.12592$ & $-0.12592$ \\
& $4.00$ & $-0.11458$ & $-0.11444$ & $-0.11445$ & $-0.11445$ & $-0.11445$ & $-0.11445$ \\
\hline
\multirow{7}{*}{$\frac{\pi}{6}$} 
& $2.00$ & $-0.15110$ & $-0.15113$ & $-0.15113$ & $-0.15113$ & $-0.15113$ & $-0.15113$ \\
& $2.50$ & $-0.13920$ & $-0.13971$ & $-0.13975$ & $-0.13974$ & $-0.13974$ & $-0.13974$ \\
& $2.90$ & $-0.12758$ & $-0.12952$ & $-0.13141$ & $-0.13019$ & $-0.13032$ & $-0.13046$ \\
& $3.00$ & $-0.12402$ & $-0.12610$ & $-0.13108$ & $-0.12705$ & $-0.12729$ & $-0.12817$ \\
& $3.10$ & $-0.12310$ & $-0.12499$ & $-0.12687$ & $-0.12564$ & $-0.12577$ & $-0.12591$ \\
& $3.50$ & $-0.11656$ & $-0.11716$ & $-0.11722$ & $-0.11721$ & $-0.11721$ & $-0.11721$ \\
& $4.00$ & $-0.10712$ & $-0.10725$ & $-0.10725$ & $-0.10725$ & $-0.10725$ & $-0.10725$ \\
\hline
\multirow{7}{*}{0} 
& $2.00$ & $-0.14420$ & $-0.14434$ & $-0.14434$ & $-0.14434$ & $-0.14434$ & $-0.14434$ \\
& $2.50$ & $-0.13010$ & $-0.13270$ & $-0.13242$ & $-0.13246$ & $-0.13245$ & $-0.13245$ \\
& $2.90$ & $-0.10950$ & $-0.13056$ & $-0.11932$ & $-0.12593$ & $-0.12186$ & $-0.12343$ \\
& $3.00$ & $-0.10054$ & $-0.13646$ & $-0.10871$ & $-0.13221$ & $-0.11144$ & $-0.12127$ \\
& $3.10$ & $-0.10552$ & $-0.12621$ & $-0.11502$ & $-0.12169$ & $-0.11754$ & $-0.11915$ \\
& $3.50$ & $-0.10822$ & $-0.11153$ & $-0.11104$ & $-0.11112$ & $-0.11111$ & $-0.11111$ \\
& $4.00$ & $-0.10153$ & $-0.10208$ & $-0.10206$ & $-0.10206$ & $-0.10206$ & $-0.10206$ \\
\hline
\end{tabular}
\caption[Convergence of gravitational potential]{Convergence of the multipole expansion of the Green function for gravitational potential ${\cal G}^\nu$ computed for a ring with radius $x'\!=\!3$ located in the equatorial plane ($\theta'\!=\!\pi/2$). Values at several latitudes $\theta$ are computed in the radial region close to the ring (from $x\!=\!2$ to $x\!=\!4$).}
\label{tab:GFNuConv}
\end{table}

\begin{table}
\centering
\begin{tabular}{|c|c|c c c c c|c|}
\hline
\multirow{2}{*}{$\theta$} & \multirow{2}{*}{$x$} & \multicolumn{5}{|c|}{number of terms added in (\ref{eq:GFOmegaDef})} & \multirow{2}{*}{exact value}\\
\cline{3-7} & & 10 & 20 & 30 & 40 & 50 & \\
\hline
\multirow{7}{*}{$\frac{\pi}{2}$}
& $2.00$ & $-0.0043591$ & $-0.0043605$ & $-0.0043605$ & $-0.0043605$ & $-0.0043605$ & $-0.0043605$ \\
& $2.50$ & $-0.0050673$ & $-0.0051175$ & $-0.0051218$ & $-0.0051222$ & $-0.0051223$ & $-0.0051223$ \\
& $2.90$ & $-0.0063084$ & $-0.0068904$ & $-0.0071424$ & $-0.0072701$ & $-0.0073402$ & $-0.0074433$ \\
& $3.00$ & $-0.0067968$ & $-0.0078323$ & $-0.0084794$ & $-0.0089513$ & $-0.0093229$ & $-\infty$ \\
& $3.10$ & $-0.0059006$ & $-0.0064596$ & $-0.0067049$ & $-0.0068309$ & $-0.0069010$ & $-0.0070079$ \\
& $3.50$ & $-0.0037462$ & $-0.0038087$ & $-0.0038162$ & $-0.0038173$ & $-0.0038174$ & $-0.0038175$ \\
& $4.00$ & $-0.0024485$ & $-0.0024550$ & $-0.0024551$ & $-0.0024551$ & $-0.0024551$ & $-0.0024551$ \\
\hline
\multirow{7}{*}{$\frac{\pi}{3}$}
& $2.00$ & $-0.0037082$ & $-0.0037087$ & $-0.0037087$ & $-0.0037087$ & $-0.0037087$ & $-0.0037087$ \\
& $2.50$ & $-0.0033726$ & $-0.0033758$ & $-0.0033779$ & $-0.0033781$ & $-0.0033782$ & $-0.0033781$ \\
& $2.90$ & $-0.0029896$ & $-0.0029065$ & $-0.0029373$ & $-0.0029732$ & $-0.0029866$ & $-0.0029767$ \\
& $3.00$ & $-0.0029039$ & $-0.0027174$ & $-0.0027375$ & $-0.0028423$ & $-0.0029251$ & $-0.0028643$ \\
& $3.10$ & $-0.0027625$ & $-0.0026820$ & $-0.0027111$ & $-0.0027462$ & $-0.0027598$ & $-0.0027498$ \\
& $3.50$ & $-0.0022901$ & $-0.0022911$ & $-0.0022943$ & $-0.0022948$ & $-0.0022948$ & $-0.0022948$ \\
& $4.00$ & $-0.0017923$ & $-0.0017936$ & $-0.0017938$ & $-0.0017938$ & $-0.0017938$ & $-0.0017938$ \\
\hline
\multirow{7}{*}{$\frac{\pi}{4}$}
& $2.00$ & $-0.0033785$ & $-0.0033783$ & $-0.0033783$ & $-0.0033783$ & $-0.0033783$ & $-0.0033783$ \\
& $2.50$ & $-0.0029105$ & $-0.0029114$ & $-0.0029139$ & $-0.0029139$ & $-0.0029139$ & $-0.0029139$ \\
& $2.90$ & $-0.0024557$ & $-0.0024287$ & $-0.0025224$ & $-0.0025304$ & $-0.0025057$ & $-0.0025097$ \\
& $3.00$ & $-0.0023163$ & $-0.0022347$ & $-0.0024540$ & $-0.0025018$ & $-0.0023813$ & $-0.0024098$ \\
& $3.10$ & $-0.0022598$ & $-0.0022333$ & $-0.0023241$ & $-0.0023322$ & $-0.0023075$ & $-0.0023116$ \\
& $3.50$ & $-0.0019376$ & $-0.0019385$ & $-0.0019426$ & $-0.0019426$ & $-0.0019425$ & $-0.0019425$ \\
& $4.00$ & $-0.0015499$ & $-0.0015498$ & $-0.0015500$ & $-0.0015500$ & $-0.0015500$ & $-0.0015500$ \\
\hline
\multirow{7}{*}{$\frac{\pi}{6}$} 
& $2.00$ & $-0.0031638$ & $-0.0031633$ & $-0.0031633$ & $-0.0031633$ & $-0.0031633$ & $-0.0031633$ \\
& $2.50$ & $-0.0026684$ & $-0.0026508$ & $-0.0026545$ & $-0.0026543$ & $-0.0026542$ & $-0.0026543$ \\
& $2.90$ & $-0.0023803$ & $-0.0021449$ & $-0.0022923$ & $-0.0022750$ & $-0.0022750$ & $-0.0022613$ \\
& $3.00$ & $-0.0023586$ & $-0.0019140$ & $-0.0022626$ & $-0.0022328$ & $-0.0020593$ & $-0.0021686$ \\
& $3.10$ & $-0.0021926$ & $-0.0019660$ & $-0.0021091$ & $-0.0020924$ & $-0.0020616$ & $-0.0020787$ \\
& $3.50$ & $-0.0017658$ & $-0.0017434$ & $-0.0017494$ & $-0.0017490$ & $-0.0017489$ & $-0.0017490$ \\
& $4.00$ & $-0.0014080$ & $-0.0014058$ & $-0.0014060$ & $-0.0014060$ & $-0.0014060$ & $-0.0014060$ \\
\hline
\multirow{7}{*}{0} 
& $2.00$ & $-0.0029876$ & $-0.0030073$ & $-0.0030070$ & $-0.0030070$ & $-0.0030070$ & $-0.0030070$ \\
& $2.50$ & $-0.0020835$ & $-0.0025531$ & $-0.0024665$ & $-0.0024810$ & $-0.0024787$ & $-0.0024790$ \\
& $2.90$ & $+0.0004811$ & $-0.0045000$ & $-0.0000854$ & $-0.0037008$ & $-0.0008545$ & $-0.0020972$ \\
& $3.00$ & $+0.0019023$ & $-0.0072430$ & $+0.0042764$ & $-0.0091957$ & $+0.0059767$ & $-0.0020094$ \\
& $3.10$ & $+0.0005373$ & $-0.0042505$ & $+0.0000482$ & $-0.0035188$ & $-0.0006733$ & $-0.0019249$ \\
& $3.50$ & $-0.0011763$ & $-0.0017360$ & $-0.0015918$ & $-0.0016257$ & $-0.0016180$ & $-0.0016194$ \\
& $4.00$ & $-0.0012397$ & $-0.0013108$ & $-0.0013062$ & $-0.0013065$ & $-0.0013064$ & $-0.0013064$ \\
\hline
\end{tabular}
\caption[Convergence of dragging]{Convergence of the multipole expansion of the Green function for dragging ${\cal G}^\omega$ computed for a ring with radius $x'\!=\!3$ located in the equatorial plane ($\theta'\!=\!\pi/2$). Values at several latitudes $\theta$ are computed in the radial region close to the ring (from $x\!=\!2$ to $x\!=\!4$).}
\label{tab:GFOmegaConv}
\end{table}

\subsection{Green function for the gravitational potential}

In order to find the closed form of the Green function (\ref{eq:GFNuDef}), one can start from a more general relation which will be proven in an accompanying paper by \cite{bib:math-part}:\footnote
{It is also shown there that the presented sum converges absolutely when $\gamma\!=\!3/2$.}
\begin{align}
  & \sum\limits_{j=0}^\infty
    2(j+\gamma)\,\frac{\Gamma^3(j+1)}{\Gamma^3(j+2\gamma)}\;
    D^{(\gamma)}_j(a)\,
    C^{(\gamma)}_j(u)\,C^{(\gamma)}_j(v)\,C^{(\gamma)}_j(w) = \nonumber \\
  &=\frac{\pi\int_{vw-\sqrt{(1-v^2)(1-w^2)}}^{vw+\sqrt{(1-v^2)(1-w^2)}}
          {}_2F_1\!\left(\frac{1}{2},1;\gamma+\frac{1}{2};\frac{(1-u^2)(1-\xi^2)}{(a-u\xi)^2}\right)
          \frac{(1-v^2-w^2-\xi^2+2vw\xi)^{\gamma-1}}{a-u\xi}\;{\rm d}\xi}
         {2^{4\gamma-2}\Gamma^4(\gamma)\Gamma(2\gamma)\,\left[(a^2-1)(1-v^2)(1-w^2)\right]^{\gamma-\frac{1}{2}}}
         \;,
  \label{eq:DCCCIntForm}
\end{align}
where $u$, $v$, $w$ $\in(-1;1)$, $a$ should be greater then 1, and $C^{(\gamma)}_j(x)$ and $D^{(\gamma)}_j(x)$ denote, respectively, Gegenbauer functions of the first and of the second kind.
Namely, $C^{(\gamma)}_j(x)$ and $D^{(\gamma)}_j(x)$ are two independent solutions of the ultraspherical differential equation 
\begin{equation}
  (1-x^2)\,\frac{{\rm d}^2 y(x)}{{\rm d}x^2}-(2\gamma+1)\,x\,\frac{{\rm d}y(x)}{{\rm d}x}
  +j(j+2\gamma)\,y(x) = 0 \,.
\end{equation}
The first solution can be written as
\begin{equation}
  C^{(\gamma)}_n(x) =
  \left|1-x^2\right|^\frac{1-2\gamma}{4}
  \frac{\sqrt{2\pi}\,\Gamma(n+2\gamma)}{2^\gamma\Gamma(\gamma)\Gamma(n+1)}\times  
     \left\{\begin{matrix}
            {\rm P}^{\frac{1}{2}-\gamma}_{-\frac{1}{2}+\gamma+n}(x)  &  {\rm when} &  |x|<1  \cr
            P^{\frac{1}{2}-\gamma}_{-\frac{1}{2}+\gamma+n}(x)  &  {\rm when} &   x \geq 1
            \end{matrix}
            \;; \right.
\end{equation}
it reduces to Gegenbauer polynomials when $n$ is a non-negative integer and to Legendre functions of the first kind (actually Legendre polynomials) for $\gamma=1/2$. 
The second solution can be written analogously,
\begin{equation}
  D^{(\gamma)}_n(x) =
  \left|1-x^2\right|^\frac{1-2\gamma}{4}
  \frac{\sqrt{2\pi}\,\Gamma(n+2\gamma)}{2^\gamma\Gamma(\gamma)\Gamma(n+1)}\times  
     \left\{\begin{matrix}
            {\rm Q}^{\frac{1}{2}-\gamma}_{-\frac{1}{2}+\gamma+n}(x)  &  {\rm when} &  |x|<1  \cr
            e^{{\rm i}\pi(\gamma-\frac{1}{2})}
            Q^{\frac{1}{2}-\gamma}_{-\frac{1}{2}+\gamma+n}(x)  &  {\rm when} &   x \geq 1
            \end{matrix}
            \;; \right.
\end{equation}
it reduces to Legendre functions of the second kind for $\gamma=1/2$ (hence how we call it). Above, we have employed
the notation ${\rm P}^{\mu}_{\nu}(x)$, $P^{\mu}_{\nu}(x)$, ${\rm Q}^{\mu}_{\nu}(x)$ and $Q^{\mu}_{\nu}(x)$, used in \cite{bib:DLMF}, chapter 14, in order to distinguish the Ferrers functions of the first and of the second kinds (i.e., Legendre functions defined on the cut $|x|\!<\!1$; written in roman) from the associated Legendre functions of the first and of the second kinds (defined on $x\!>\!1$; written in italic).

One more remark to the formula (\ref{eq:DCCCIntForm}) is at place, namely that it requires $u$ to lie within $(-1;1)$, while for a ring outside of the horizon actually $u\!>\!1$. However, a function of complex variable has a unique extension (up to a Riemann folding), so the solution valid for $u \in (-1;1)$, when extended into the complex plane, should also yield a (unique) solution of the respective differential equation with different boundary conditions.

The relation (\ref{eq:DCCCIntForm}) provides ${\cal G}^\nu(x,\theta,x',\theta')$ (up to a multiplication factor) if putting $\gamma\!=\!1/2$, $a\!=\!\max(x,x')$, $u\!=\!\min(x,x')$, $v\!=\!\cos\theta$ and $w\!=\!\cos\theta'$.
Namely, as will be shown in \cite{bib:math-part} (the result follows from \citealt{bib:Baranov2006}),
\begin{equation}
  \sum\limits_{j=0}^\infty (2j+1)\,Q_j(a)\,P_j(u)\,P_j(v)\,P_j(w)
  = \frac{2\,K\!\left[\sqrt{\frac{4\;\sqrt{(a^2-1)(u^2-1)(1-v^2)(1-w^2)}}
                              {(au-vw)^2-\left(\sqrt{(a^2-1)(u^2-1)}-\sqrt{(1-v^2)(1-w^2)}\right)^2}}\right]}
         {\pi \left[(au-vw)^2-\left(\sqrt{(a^2-1)(u^2-1)}-\sqrt{(1-v^2)(1-w^2)}\right)^2\right]^{1/2}} \;, 
\end{equation}
hence, comparing this with (\ref{eq:GFNuDef}), we get
\begin{equation}
  {\cal G}^\nu(x,\theta,x',\theta')
  =-\frac{K\!\left[\sqrt{\frac{4\;\sqrt{(x^2-1)(x'^2-1)}\,\sin\theta\sin\theta'}
                              {(xx'-\cos\theta\cos\theta')^2
                               -\left(\sqrt{(x^2-1)(x'^2-1)}-\sin\theta\sin\theta'\right)^2}}\right]}
         {\pi \left[(xx'-\cos\theta\cos\theta')^2
                    -\left(\sqrt{(x^2-1)(x'^2-1)}-\sin\theta\sin\theta'\right)^2\right]^{1/2}} \;.
  \label{eq:GFNu}
\end{equation}
This is actually an expected result: the Green function corresponds to the potential due to a thin-ring source (situated at $x'$, $\theta'$), which is familiar to be given by the complete elliptic integral $K(k)$ (within general relativity, such a source is known as the Bach-Weyl ring, see e.g. \citealt{bib:Semer2016}). Note in passing that equation (\ref{eq:EFE2}) is quadratic in dragging ($\omega$) and the background metric is static (Schwarzschild solution), so dragging has to be proportional to the perturbation ($\lambda$), i.e. the correction from the $(\nabla\omega)^2$ term is at least of order $\lambda^2$ and does not contribute in the linear order (in higher orders, it behaves like a source term). Without dragging, equation (\ref{eq:EFE2}) is the same as in the static case (or even a Newtonian one) and one reaches the above result (up to a coordinate transformation, see (\ref{eq:CoordinateTransform})).

\subsection{Green function for dragging}

Now we proceed to the second Green function (\ref{eq:GFOmegaDef}).
Considering relations
\begin{align}
  & F_j(x) = \frac{12\,(-1)^j}{j(j+1)^2(j+2)^2(j+3)}\;\hat{{\rm O}}_x C^{(3/2)}_j(x) \,,
    \qquad
    G_j(x) = \frac{(-1)^j}{12}\;\hat{{\rm O}}_x D^{(3/2)}_j(x) \,, \label{eq:CDDifForm} \\
  & {\rm where} \quad
    \hat{{\rm O}}_x := (x-1)\,\frac{{\rm d}^2}{{\rm d}x^2}\,(x-1)
                     = \frac{{\rm d}}{{\rm d}x}\,(x-1)^2\frac{{\rm d}}{{\rm d}x}\,,
\end{align}
we can rewrite (\ref{eq:GFOmegaDef}) as
\begin{align}
  {\cal G}^\omega
  &= \Delta - 
     \sum\limits_{j=1}^\infty \frac{j+\frac{3}{2}}{j(j+1)^3(j+2)^3(j+3)}\;
     \hat{{\rm O}}_{x_<}\,C^{(3/2)}_j(x_<)\,
     \hat{{\rm O}}_{x_>}\,D^{(3/2)}_j(x_>)\,C^{(3/2)}_j(\cos\theta)\,C^{(3/2)}_j(\cos\theta') = \nonumber \\
  &= \Delta -
     \hat{{\rm O}}_{x_>}\,\frac{{\rm d}}{{\rm d}x_<}\,(x_< -1)^2
     \sum\limits_{j=1}^\infty \frac{j+\frac{3}{2}}{j(j+1)^3(j+2)^3(j+3)}\;
     D^{(3/2)}_j(x_>)\,\frac{{\rm d}}{{\rm d}x_<}\,
     C^{(3/2)}_j(x_<)\,C^{(3/2)}_j(\cos\theta)\,C^{(3/2)}_j(\cos\theta') = \nonumber\\
  &= \widetilde\Delta -
     \hat{{\rm O}}_{x_>}\,\frac{{\rm d}}{{\rm d}x_<}\,(x_< +1)^{-2}
     \sum\limits_{j=0}^\infty \frac{j+\frac{3}{2}}{(j+1)^3(j+2)^3}
     \int\limits_{1}^{x_<}(\zeta^2-1)\,
     D^{(3/2)}_j(x_>)\,C^{(3/2)}_j(\zeta)\,C^{(3/2)}_j(\cos\theta)\,C^{(3/2)}_j(\cos\theta')\;
     {\rm d}\zeta = \nonumber \\
  &= \widetilde\Delta -
     \hat{{\rm O}}_{x_>}\,\frac{{\rm d}}{{\rm d} x_<}
     \int\limits_{1}^{x_<}\frac{\zeta^2-1}{2(x_< +1)^2}
     \sum\limits_{j=0}^\infty \frac{2j+3}{(j+1)^3(j+2)^3}\;
     D^{(3/2)}_j(x_>)\,C^{(3/2)}_j(\zeta)\,C^{(3/2)}_j(\cos\theta)\,C^{(3/2)}_j(\cos\theta')\;
     {\rm d}\zeta = \label{eq:GNuIntForm} \\
  &= \widetilde\Delta -
     \hat{{\rm O}}_{x_>}\,\frac{{\rm d}}{{\rm d} x_<}
     \int\limits_{1}^{x_<}\!\frac{\zeta^2-1}{2\pi(x_<\!+\!1)^2(x_>^2\!-\!1)\,\sin^2\theta\sin^2\theta'} \!
     \int\limits_{\cos(\theta+\theta')}^{\cos(\theta-\theta')} \!
     \frac{\sqrt{[\cos(\theta-\theta')-\xi][\xi-\cos(\theta+\theta')]}}
          {x_>-\zeta\xi+\sqrt{(x_>\!-\!\zeta\xi)^2-(1\!-\!\zeta^2)(1\!-\!\xi^2)}}\;
     {\rm d}\xi {\rm d}\zeta \,, \nonumber
\end{align}
where
\begin{align}
   \Delta &:= -\frac{3}{4}\,F_0(x_<)\,G_0(x_>)\,C^{(3/2)}_0(\cos\theta)\,C^{(3/2)}_0(\cos\theta')
            = -\frac{1}{4\,(x_>+1)^3} \;, \\
   \widetilde\Delta &:= \Delta +
                    \hat{{\rm O}}_{x_>}\,\frac{{\rm d}}{{\rm d} x_<}\,\frac{3}{16\,(x_< +1)^2}
                    \int\limits_{1}^{x_<} (\zeta^2-1)\,
                    D^{(3/2)}_0(x_>)\,C^{(3/2)}_0(\zeta)\,C^{(3/2)}_0(\cos\theta)\,C^{(3/2)}_0(\cos\theta')\;
                    {\rm d}\zeta = \nonumber \\
                & = -\frac{2}{(x+1)^3(x'+1)^3} \;. \label{eq:DeltaP}
\end{align}
We have deliberately switched the order of summation, integration and differentiation, which might in fact be an issue, but we will later check that the result is really a Green function of the original equation (\ref{eq:PertEFE3}) with given boundary conditions.
Note that the above result is simple thanks to its symmetry with respect to $x_<\!\leftrightarrow x_>$, and also due to the relation
\begin{equation}
  \int\limits_1^x (z^2-1)\,C^{(3/2)}_n(z)\;{\rm d}z =
  \frac{(x^2-1)^2}{n(n+3)}\,\frac{{\rm d}C^{(3/2)}_n(x)}{{\rm d} x}
\end{equation}
which is a direct consequence of the ultraspherical differential equation.

Calculation of the integral (\ref{eq:GNuIntForm}) leads to logarithms or elliptic functions, but, ``miraculously'', when one applies the operator $\hat{{\rm O}}_{x_>}$ first and only then integrates by $\zeta$, such difficulties do not occur, because
\begin{equation}
  \frac{{\rm d}}{{\rm d} x_<}
  \int\limits_{1}^{x_<} \hat{{\rm O}}_{x_>}
  \frac{(\zeta^2-1)\;{\rm d}\zeta}
       {2\pi(x_<\!+\!1)^2(x_>^2\!-\!1)\,\sin^2\theta \sin^2\theta'
        \left[x_> -\zeta\xi+\sqrt{(x_>\!-\!\zeta\xi)^2-(1\!-\!\zeta^2)(1\!-\!\xi^2)}\right]}
  = \sum\limits_{k=0}^5 {\cal P}_k(x',\theta')\,\tilde{I}_k(x',\xi) \,, \label{eq:GF_Omega_InnerInt}
\end{equation}
where we have denoted
\begin{align}
  {\cal P}_ 0(x',\theta') &:= \frac{-2}{\pi(x+1)^3(x'+1)^3\sin^2\theta\sin^2\theta'} \,,
     & \tilde{I}_0(x',\xi) &:= 1 \,, \nonumber \\ 
  {\cal P}_ 1(x',\theta') &:= \frac{-(x-1)(x'-1)}{\pi(x+1)^3 (x'+1)^3\sin^2\theta\sin^2\theta'} \,,
     & \tilde{I}_1(x',\xi) &:= \frac{1}{\xi+1} \,, \nonumber \\
  {\cal P}_ 2(x',\theta') &:= \frac{2xx'+(x-1)(x'-1)}{\pi(x+1)^3(x'+1)^3\sin^2\theta\sin^2\theta'} \,,
     & \tilde{I}_2(x',\xi) &:= \frac{1}{\sqrt{x^2+x'^2+\xi^2-1-2xx'\xi}} \,, \nonumber \\
  {\cal P}_ 3(x',\theta') &:= \frac{-2}{\pi(x+1)^3(x'+1)^3\sin^2\theta\sin^2\theta'} \,,
     & \tilde{I}_3(x',\xi) &:= \frac{\xi}{\sqrt{x^2+x'^2+\xi^2-1-2xx'\xi}} \,, \nonumber \\
  {\cal P}_ 4(x',\theta') &:= \frac{(x'-1)(x-1)(x'+x)}{\pi(x+1)^3(x'+1)^3\sin^2\theta\sin^2\theta'} \,,
     & \tilde{I}_4(x',\xi) &:= \frac{1}{(\xi+1)\sqrt{x^2+x'^2+\xi^2-1-2xx'\xi}} \,, \nonumber \\
  {\cal P}_ 5(x',\theta') &:= \frac{(x-1)(x'-1)}{2\pi(x+1)(x'+1)\sin^2\theta\sin^2\theta'} \,,
     & \tilde{I}_5(x',\xi) &:= \frac{1}{\left(x^2+x'^2+\xi^2-1-2xx'\xi\right)^{3/2}} \,.
  \label{eq:IImplicit}
\end{align}
Therefore, we can finally conclude that
\begin{align}
  {\cal G}^\omega(x,\theta,x',\theta')
  &= \widetilde\Delta(x,x') - \!\!
     \int\limits_{\cos(\theta+\theta')}^{\cos(\theta-\theta')}\!\!
     \sqrt{[\cos(\theta\!-\!\theta')-\xi][\xi-\cos(\theta\!+\!\theta')]}\;
     \sum\limits_{k=0}^5 {\cal P}_k(x',\theta')\,\tilde{I}_k(x',\xi)\;{\rm d}\xi = \nonumber \\
  &= \widetilde\Delta(x,x') - \sum\limits_{k=0}^5 {\cal P}_ k(x',\theta')\,I_k(x',\theta') \,,
\end{align}
where
\begin{align}
  I_0(x',\theta') &:= \frac{\pi}{2}\sin^2\theta\,\sin^2\theta' \,, \nonumber \\
  I_1(x',\theta') &:= \pi\,(1-|\cos\theta|)(1-|\cos\theta'|) \,, \nonumber \\
  I_2(x',\theta') &:= \sqrt{\frac{a_{31}}{a_{42}}}
                      \left[-a_{41}K(k)-a_{42}E(k)
                            +a_{41}\!\left(1+\frac{a_{42}}{a_{31}}\right)
                             \Pi\!\left(\!-\frac{a_{43}}{a_{31}},k\right)\right], \nonumber \\
  I_3(x',\theta') &:= \frac{a_{41}}{4}\,\sqrt{\frac{a_{31}}{a_{42}}}
                      \left[-(s_{21}\!+\!s_{43}\!+\!2a_{21})K(k)
                            +\frac{a_{42}}{a_{41}}\,(s_{43}\!-\!3s_{21})E(k)
                            +\frac{a_{21}^2\!-\!a_{43}^2\!+\!2s_{21}^2\!-\!2s_{21}s_{43}}{a_{31}}\;
                             \Pi\!\left(\!-\frac{a_{43}}{a_{31}},k\right)\right], \nonumber \\
  I_4(x',\theta') &:= \frac{2a_{41}}{\sqrt{a_{31}a_{42}}}
                      \left[-\frac{K(k)}{a_1+1}+\Pi\!\left(\!-\frac{a_{43}}{a_{31}},k\right)
                            -\frac{a_3+1}{a_1+1}\,
                             \Pi\!\left(\!-\frac{(a_1+1)a_{43}}{(a_4+1)a_{31}},k\right)\right], \nonumber \\
  I_5(x',\theta') &:= \frac{2\,\sqrt{a_{31}a_{42}}}{a_{21}^2}
                      \left[(2-k^2)K(k)-2E(k)\right],  \label{eq:IExplicit}
\end{align}
with
\begin{equation}  \label{eq:a_i}
  a_1 := xx'+\sqrt{(x^2-1)(x'^2-1)} \,, \quad
  a_2 := xx'-\sqrt{(x^2-1)(x'^2-1)} \,, \quad
  a_3 := \cos(\theta-\theta'), \quad
  a_4 := \cos(\theta+\theta')
\end{equation}
and
\begin{equation}  \label{eq:a_rs,k}
  a_{rs} := a_s - a_r \,, \qquad
  s_{rs} := a_s + a_r \,, \qquad
  k := \sqrt{\frac{a_{21}a_{43}}{a_{31}a_{42}}} \;.
\end{equation}

On the horizon the above Green function simplifies considerably,
\begin{equation}
  {\cal G}^{\omega}(x\!=\!1,\theta,x',\theta') = -\frac{1}{4(x'+1)^3} \;.
\end{equation}
Note that this is independent of the position on the horizon (i.e. of the angle $\theta$), consistently with the well known ``rigid rotation'' of stationary horizons.

\section{Disc solution}
\label{ch:Disc}

Perturbation equations can be solved by integrating the Green functions over the source mass distribution. Due to the presence of (complete) elliptic integrals in the above concise expression of the Green functions, a numerical integration has to be employed in general. Integration is of course simpler for thin sources (surface distribution) than for extended ones. Below we consider the case of a stationary and axially symmetric thin disc, encircling the central black hole between some two finite radii $x_{\rm in} \leq x\equiv\frac{r}{M}+\frac{M}{4r}\leq x_{\rm out}$ in a concentric manner (thus lying in the equatorial plane). We will show how to obtain explicit results for the linear perturbation of the metric functions and for the corresponding basic characteristics (density, pressure and velocity distribution) of the source.

\subsection{Gravitational potential}
\label{grav-potential}

First, let us find the linear perturbation of potential, $\nu_1$.
It is advantageous to choose $B\!=\!1$ here, since this makes equation (\ref{eq:PertEFE2}) the Laplace/Poisson equation, so one can use its solutions known from the classical field theory. Let us denote by $t$, $\phi$, $\tilde{\rho}$ and $\tilde{z}$ the coordinates corresponding to this choice (time $t$ and azimuth $\phi$ do not depend on the choice of $B$). Their counterparts ($\rho$, $z$) corresponding to $B\!=\!1\!-\!\frac{M^2}{4r^2}\,$ are related by
\begin{align}
  \tilde{\rho} &= \frac{\rho\,(M^2-4\rho^2-4z^2)}{4(\rho^2+z^2)}
                = \frac{(4r^2-M^2)\sin\theta}{4r}
                = M\sqrt{x^2-1}\,\sin\theta \,, \nonumber \\
  \tilde{z} &= \frac{z\,(M^2+4\rho^2+4z^2)}{4(\rho^2+z^2)}
             = \frac{(4r^2+M^2)\cos\theta}{4r}
             = Mx\cos\theta \,, \label{eq:CoordinateTransform} \\
  \tilde{\zeta} &= \zeta-\frac{1}{2}\ln\!\left[\frac{(4\rho^2\!+4z^2\!+\!M^2)^2-16M^2 z^2}{16(\rho^2+z^2)^2}\right]
                 = \zeta-\frac{1}{2}\ln\!\left[\frac{(4r^2\!+\!M^2)^2-16r^2 M^2\cos^2\theta}{16r^4}\right]
                 = \zeta-\frac{1}{2}\ln\frac{4\,(x^2\!-\cos^2\theta)}{\left(x+\sqrt{x^2\!-\!1}\right)^2}
                   \nonumber
\end{align}
(to recall the function $\zeta$, let us remind that $g_{\rho\rho}=g_{rr}=e^{2\zeta-2\nu}$).

Of the known disc solutions of the Laplace equation, we will choose the one describing a disc which extends from the origin ($\tilde{\rho}\!=\!\tilde{z}\!=\!0$) to some finite equatorial radius $x'$ (or $\tilde{\rho}'$ when expressed in the Weyl radius corresponding to the $B\!=\!1$ choice) and whose Newtonian surface density $S$ is constant in the unperturbed $B\!=\!1$ coordinate system. Such a solution was obtained by \cite{bib:LassBlitzer1983},
\begin{align}
  & V(x';x,\theta)= \nonumber \\
  & = 2\pi S|\tilde{z}|\,H(\tilde{\rho}'\!-\!\tilde{\rho})
      -\frac{2S}{\sqrt{\tilde{z}^2+(\tilde{\rho}'\!+\!\tilde{\rho})^2}}
       \left\{\left[\tilde{z}^2+(\tilde{\rho}'\!+\!\tilde{\rho})^2\right]E(k)
              +(\tilde{\rho}'^2\!-\!\tilde{\rho}^2)\,K(k)
              +\tilde{z}^2\,\frac{\tilde{\rho}'\!-\!\tilde{\rho}}{\tilde{\rho}'\!+\!\tilde{\rho}}\;
               \Pi\!\left[\frac{4\tilde{\rho}'\tilde{\rho}}{(\tilde{\rho}'\!+\!\tilde{\rho})^2},k\right]
              \right\} = \nonumber \\
  & = 2\pi MSx|\cos\theta|\,H(x'^2\!-\!\cos^2\theta\!-\!x^2\sin^2\theta)- {} \nonumber \\
  & \quad
    -2MS\;
     \frac{a_{31}a_{42}\,E(k)
           +(x'^2\!-\cos^2\theta-x^2\sin^2\theta)\,K(k)
           +\frac{(a_{41}a_{32}-x^2\cos^2\theta)\,x^2\cos^2\theta}{x'^2-\cos^2\theta-x^2\sin^2\theta}\;
            \Pi\!\left(\frac{2a_{21}\sin\theta}{a_{31}a_{42}-x^2\cos^2\theta}\,,k\right)}
          {\sqrt{a_{31}a_{42}}} \;,  \label{LassBlitzer-potential}
\end{align}
where $H(x)$ stands for the Heaviside function, and the $a_{rs}(x')$ and $k(x')$ symbols are given by (\ref{eq:a_i}) and (\ref{eq:a_rs,k}) evaluated at $\theta'\!=\!\pi/2$,
\begin{align}
  &
  a_1(x') = xx'+\sqrt{(x^2-1)(x'^2-1)} \,, \quad
  a_2(x') = xx'-\sqrt{(x^2-1)(x'^2-1)} \,, \quad
  a_3(x') = \sin\theta, \quad
  a_4(x') = -\sin\theta, \nonumber \\
  &
  a_{rs}(x') = a_s(x')-a_r(x'), \qquad
  k^2(x') = \frac{2a_{21}\sin\theta}{a_{31}a_{42}}
          = \frac{4\,\sqrt{(x^2-1)(x'^2-1)}\;\sin\theta}
                 {x^2 x'^2-\left[\sqrt{(x^2\!-\!1)(x'^2\!-\!1)}-\sin\theta\right]^2}
          = \frac{4\tilde{\rho}'\tilde{\rho}}{(\tilde{\rho}'+\tilde{\rho})^2+\tilde{z}^2} \;.
  \label{eq:AXPrimeDef}
\end{align}

The potential of a disc extending between two finite radii $x_{\rm in}\!\leq\!x\!\leq\!x_{\rm out}$ (which we are interested in) is obtained by subtraction of two potentials (\ref{LassBlitzer-potential}) (with the same density $S$) with outer radii at $x_{\rm out}$ and $x_{\rm in}$,
\begin{equation}  \label{nu1}
  \nu_{1}(x,\theta)=V(x'\!=\!x_{\rm out};x,\theta)-V(x'\!=\!x_{\rm in};x,\theta) \,.
\end{equation}

\subsection{Dragging}

In order to find the perturbation of dragging angular velocity, one has to integrate (\ref{eq:GNuIntForm}) over the ``density'' of the disc $W(x)$ (here assumed to be constant, $W$). Since, as already noticed, (\ref{eq:GNuIntForm}) is symmetric with respect to the exchange $x_<\!\leftrightarrow x_>$, we can substitute $x_<$ by $x'$ and $x_>$ by $x$ without changing the result.
\begin{align}
  &\omega_1(x, \theta)
   = -\int\limits_{x_{\rm in}}^{x_{\rm out}} W\,{\cal G}^\omega(x,\omega,x',\pi/2)\,{\rm d}x'= \nonumber \\
  &= -\int\limits_{x_{\rm in}}^{x_{\rm out}} W\,\widetilde\Delta(x,x')\,{\rm d} x'
     +\int\limits_{x_{\rm in}}^{x_{\rm out}} W\;\hat{{\rm O}}_{x}\,\frac{{\rm d}}{{\rm d}x'}
      \int\limits_{1}^{x'}\frac{\zeta^2-1}{\pi(x'+1)^2\sin^2\theta\left(x^2-1\right)}
      \int\limits_{-\sin\theta}^{\sin\theta}
      \frac{\sqrt{\sin^2\theta-\xi^2}\;\;{\rm d}\xi\,{\rm d}\zeta\,{\rm d}x'}
           {x-\zeta\xi+\sqrt{(x-\zeta\xi)^2-(1-\zeta^2)(1-\xi^2)}}= \nonumber \\
  &= \frac{(x_{\rm out}-x_{\rm in})(x_{\rm out}+x_{\rm in}+2)\,W}{(x+1)^3(x_{\rm in}+1)^2(x_{\rm out}+1)^2}
     + W \!\!\!
       \int\limits_{-\sin\theta}^{\sin\theta}
       \left[\frac{\hat{{\rm O}}_{x}}{2\pi\sin^2\theta}
       \int\limits_{1}^{x'}
       \frac{\sqrt{\sin^2\theta-\xi^2}}
            {x\!-\!\zeta\xi\!+\!\sqrt{(x\!-\!\zeta\xi)^2\!-\!(1\!-\!\zeta^2)(1\!-\!\xi^2)}}\;
       \frac{(\zeta^2-1)\;\,{\rm d}\zeta}{(x'\!+\!1)^2(x^2\!-\!1)}\right]_{x'=x_{\rm in}}^{x_{\rm out}} \!\!\!
       {\rm d}\xi = \nonumber \\
  &= \frac{(x_{\rm out}-x_{\rm in})(x_{\rm out}+x_{\rm in}+2)\,W}{(x+1)^3(x_{\rm in}+1)^2(x_{\rm out}+1)^2}
     + W \!\!\!
       \int\limits_{-\sin\theta}^{\sin\theta}
       \sqrt{\sin^2\theta-\xi^2}\;\sum\limits_{j=0}^7
       \left[Q_j(x_{\rm out})\,\tilde{I}_j(x_{\rm out},\xi)-Q_j(x_{\rm in})\,\tilde{I}_j(x_{\rm in},\xi)\right]
       {\rm d}\xi=  \nonumber \\
  &= \frac{(x_{\rm out}-x_{\rm in})(x_{\rm out}+x_{\rm in}+2)\,W}{(x+1)^3(x_{\rm in}+1)^2(x_{\rm out}+1)^2}
     + W\;\sum\limits_{j=0}^7
          \left[Q_j(x_{\rm out},\theta)\,I_j(x_{\rm out},\theta)
                -Q_j(x_{\rm in},\theta)\,I_j(x_{\rm in},\theta)\right],
  \label{LassBlitzer-omega_1}
\end{align}
where $x_{\rm in}$ and $x_{\rm out}$ mark, again, the inner and outer rims of the disc (in the coordinate $x$), $\tilde{I}_0$, \dots , $\tilde{I}_5$ are defined by (\ref{eq:IImplicit}),
\begin{align}
  \tilde{I}_6(x',\xi) &= \frac{1}{\xi-1} \,, &
  \tilde{I}_7(x',\xi) &= \frac{1}{(\xi-1)\sqrt{x^2+x'^2+\xi^2-1-2xx'\xi}} \,, \nonumber \\
  Q_0(x') &= \frac{1}{\pi\,(x+1)^3(x'+1)^2\sin^2\theta} \,, &
  Q_1(x') &= \frac{(1-x)(x'-1)^2}{4\pi\,(x+1)^3(x'+1)^2\sin^2\theta} \,, \nonumber \\
  Q_2(x') &= \frac{x'(1-2x)}{\pi\,(x+1)^3(x'+1)^2\sin^2\theta} \,, &
  Q_3(x') &= \frac{1}{\pi\,(x+1)^3(x'+1)^2\sin^2\theta} \,, \nonumber \\
  Q_4(x') &= \frac{(x+x')(x-1)(x'-1)^2}{4\pi\,(x+1)^3(x'+1)^2\sin^2\theta} \,, &
  Q_5(x') &= 0 \,, \nonumber \\
  Q_6(x') &= \frac{(1-x)}{4\pi\,(x+1)^3\sin^2\theta} \,, & 
  Q_7(x') &= \frac{(1-x)(x'-x)}{4\pi\,(x+1)^3\sin^2\theta} \,,
\end{align}
and $I_k(x')$ correspond to (\ref{eq:IExplicit}) taken at $\theta'\!=\!\pi/2$,
\begin{align}
  I_0(x') &= \frac{\pi}{2}\sin^2\theta \,, \nonumber \\
  I_1(x') &= \pi\,(1-|\cos\theta|) \,, \nonumber \\
  I_2(x') &= \sqrt{\frac{a_{31}}{a_{42}}}
             \left[-a_{41}K(k)-a_{42}E(k)
                   +2xx'\,\frac{a_{41}}{a_{31}}\;\Pi\!\left(\!-\frac{2\sin\theta}{a_{31}},k\right)
                  \right], \nonumber \\
  I_3(x') &= a_{41}\,\sqrt{\frac{a_{31}}{a_{42}}}
             \left[\frac{a_2\!-3a_1}{4}\,K(k)-\frac{3a_{42}}{2a_{41}}\,xx' E(k)
                   +\frac{x^2 x'^2\!+\cos^2\theta-(x\!-\!x')^2}{a_{31}}\;
                    \Pi\!\left(\!-\frac{2\sin\theta}{a_{31}},k\right)\!\right], \nonumber \\
  I_4(x') & = \frac{2a_{41}}{\sqrt{a_{31}a_{42}}}
              \left\{-\frac{a_{31}}{a_1+1}\,K(k)
                     +\Pi\!\left(\!-\frac{2\sin\theta}{a_{31}},k\right)
                     -\frac{1+\sin\theta}{a_1+1}\;
                      \Pi\!\left[-\frac{2\sin\theta\,(a_1+1)}{a_{31}(1-\sin\theta)},k\right]
                    \right\}, \nonumber \\
  I_5(x') & = \frac{\sqrt{a_{31}a_{42}}}{2(x^2-1)(x'^2-1)}
              \left[(2-k^2)\,K(k)-2E(k)\right], \nonumber \\
  I_6(x') & = \pi\,(|\cos\theta|-1) \,, \nonumber \\
  I_7(x') & = \frac{2a_{41}}{\sqrt{a_{31}a_{42}}}
              \left\{-\frac{a_{31}}{a_1-1}\,K(k)
                     +\Pi\!\left(\!-\frac{2\sin\theta}{a_{31}},k\right)
                     +\frac{1-\sin\theta}{a_1-1}\;
                      \Pi\!\left[\frac{2\sin\theta\,(a_1-1)}{a_{31}(1+\sin\theta)},k\right]
                    \right\},
\end{align}
where the argument $(x')$ of $a_i$, $a_{rs}$ and $k$ (given by (\ref{eq:AXPrimeDef})) has been omitted.

\section{Behaviour of the solution at significant locations}

Regarding the axial and reflectional symmetry, one naturally asks whether and how the solution simplifies on the axis and in the equatorial plane.
On the axis, $\sin\theta\!=\!0$ and $\cos^2\theta\!=\!1$, so, in the potential (\ref{LassBlitzer-potential}), $a_3\!=\!0$, $a_4\!=\!0$, $k\!=\!0$, so all the elliptic integrals there reduce to $\pi/2$ and
\[a_{31}a_{42}=a_{41}a_{32}=a_1 a_2=x'^2+x^2-1 \,.\]
Finally, considering that the disc has to lie above the horizon, i.e. at $x'\!>\!1$, we have $H(x'^2\!-\!1)=1$, so the first term of (\ref{LassBlitzer-potential}) is the same for both $x'\!=\!x_{\rm in}$ and $x'\!=\!x_{\rm out}$ (and thus cancels out), and hence
\begin{equation}
  \nu_1(x,\theta\!=\!0)=
  -2\pi MS\left(\sqrt{x_{\rm out}^2+x^2-1}-\sqrt{x_{\rm in}^2+x^2-1}\right).
\end{equation}
Also the equatorial form of the potential (\ref{LassBlitzer-potential}) is somewhat simpler, namely the first term and the one with the $\Pi$ integral vanish, so one is left with
\begin{equation}
  \nu_1(x,\theta\!=\!\pi/2)=
  -2MS\left[\frac{(a_1-1)(a_2+1)\,E(k)+(x'^2-x^2)\,K(k)}{\sqrt{(a_1-1)(a_2+1)}}\right]_{x'=x_{\rm in}}^{x'=x_{\rm out}}.
\end{equation}
At radial infinity ($r\!\rightarrow\!\infty$) the disc potential falls off as
\begin{equation}  \label{nu1,r=infty}
  \nu_1(r\!\rightarrow\!\infty)\propto
  -\frac{\pi S}{r}\left[r_{\rm out}^2-r_{\rm in}^2
                        -\frac{M^4}{16}\left(\frac{1}{r_{\rm in}^2}-\frac{1}{r_{\rm out}^2}\right)\right]
\end{equation}
and at the horizon ($x\!=\!1$) it assumes the value
\begin{equation}
  \nu_1(x\!=\!1,\theta)=-2\pi MS\left(\sqrt{x_{\rm out}^2-\sin^2\theta}-\sqrt{x_{\rm in}^2-\sin^2\theta}\right).
\end{equation}

The formula for the dragging perturbation $\omega_1(x,\theta)$ is more tricky and does not reduce so much on the axis and in the equatorial plane (however, figure \ref{fig:dragging} shows that it behaves reasonably). Very simple limits can still be obtained at radial infinity where it falls off as
\begin{equation}  \label{omega1,r=infty}
  \omega_1(x\!\rightarrow\!\infty)\propto \frac{W}{4}\,\frac{x_{\rm out}-x_{\rm in}}{x^3}
  \qquad \Longrightarrow \qquad
  \omega_1(r\!\rightarrow\!\infty)\propto
         \frac{WM^2}{4r^3}\left[r_{\rm out}-r_{\rm in}+
                                \frac{M^2}{4}\left(\frac{1}{r_{\rm out}}-\frac{1}{r_{\rm in}}\right)\right],
\end{equation}
and on the horizon where it is everywhere the same (independent of $\theta$) and having its sign given by $W$,
\begin{equation}  \label{omega1,H}
  \omega_1(x\!=\!1)\equiv\omega_{\rm H}
                   =\frac{W}{8}\,\frac{(x_{\rm out}-x_{\rm in})(x_{\rm out}+x_{\rm in}+2)}
                                      {(x_{\rm out}+1)^2(x_{\rm in}+1)^2} \;.
\end{equation}

\section{Parameters of the disc source}

In order to calculate the physical characteristics of the disc, let us realize that all the densities and one-stream pressure $(\sigma,P,\sigma_+,\sigma_-)$ are themselves {\em small}, namely of linear perturbation order (see equation (\ref{eq:Density})) and that the geodesic orbital velocities (\ref{Omega_pm}) are given by their unperturbed Schwarzschild values ($\pm\Omega_0$, to which corresponds $v_+\!=\!-v_-\!=\!v_0\!=\!2\sqrt{Mr}/(2r\!-\!M)$) plus terms $O(\omega)$, where $\omega$ is (of course) linearly small. Consequently, up to the linear order, relations (\ref{nuz,jump})--(\ref{zetaz,jump}) between normal jumps of the metric gradients across the disc and its physical parameters reduce to
\begin{align}
  \nu_{1,z}(z\!=\!0^+)
    &= 2\pi\left(S^\phi_\phi-S^t_t\right)
     = 2\pi (\sigma+P)\,\frac{1+v^2}{1-v^2}
     = 2\pi(\sigma_++\sigma_-)\,\frac{1+v_0^2}{1-v_0^2} \;,
     \label{nuz,jump,1} \\
  \omega_{1,z}(z\!=\!0^+)
    &=-\frac{8\pi\,S^t_\phi}{B^2\rho^2 e^{-4\nu_0}}
     = -8\pi (\sigma+P)\,\frac{\Omega}{1-v^2}
     = -8\pi(\sigma_+-\sigma_-)\,\frac{\Omega_0}{1-v_0^2} \;,
     \label{omegaz,jump,1} \\
  \zeta_{1,z}(z\!=\!0^+)
    &= 4\pi S^\phi_\phi
     = 4\pi\,\frac{\sigma v^2+P}{1-v^2}
     = 4\pi(\sigma_++\sigma_-)\,\frac{v_0^2}{1-v_0^2} \;.
     \label{zetaz,jump,1}
\end{align}

\begin{figure}[ht]
\includegraphics[width=\textwidth]{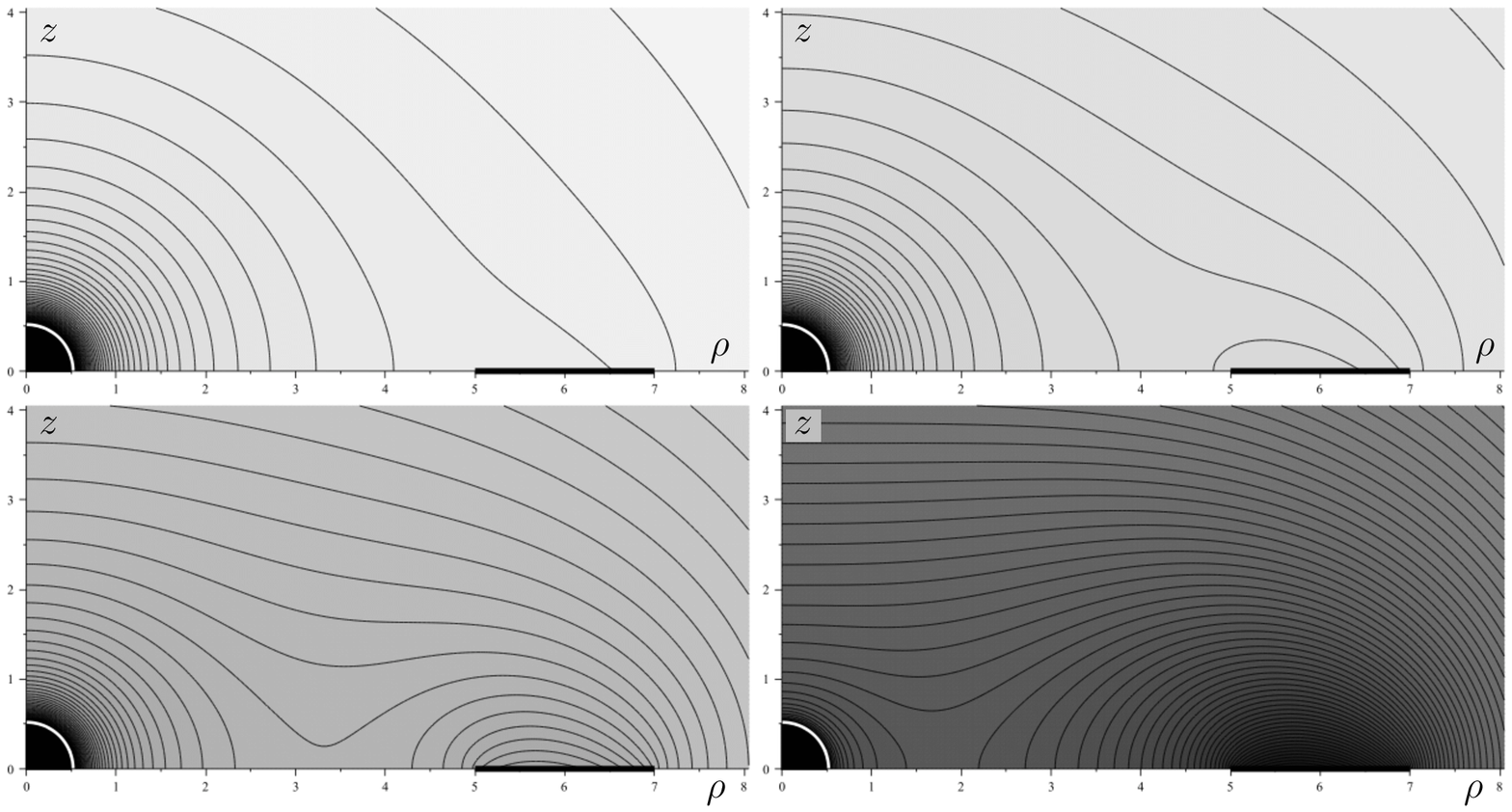}
\caption
{Meridional-plane contours of the gravitational potential $\nu$, given by sum of the Schwarzschild expression and the contribution from the disc. The four examples shown represent an equatorial disc stretching from $\rho\!=\!5M$ to $\rho\!=\!7M$ (it is indicated by a thick black line), with potential (\ref{LassBlitzer-potential}) scaled by $S\!=\!0.01$ (top left), $S\!=\!0.026$ (top right), $S\!=\!0.1$ (bottom left) and $S\!=\!1.0$ (bottom right) (such a series corresponds to a more and more massive disc). The potential is everywhere negative, with light/dark shading indicating shallow/deeper values (the potential diverges to $-\infty$ at the horizon, while the ``weakest'' levels reached at top right corners of the plots amount to $-0.20$ at top left, to $-0.34$ at top right, to $-1.00$ at bottom left and to $-8.96$ at bottom right); the black-hole horizon (at $\rho^2+z^2=M^2/4$) is represented by the white quarter-circle. Both axes are given in the units of $M$.}
\label{fig:potential}
\end{figure}

\begin{figure}[ht]
\begin{center}
\includegraphics[width=0.94\textwidth]{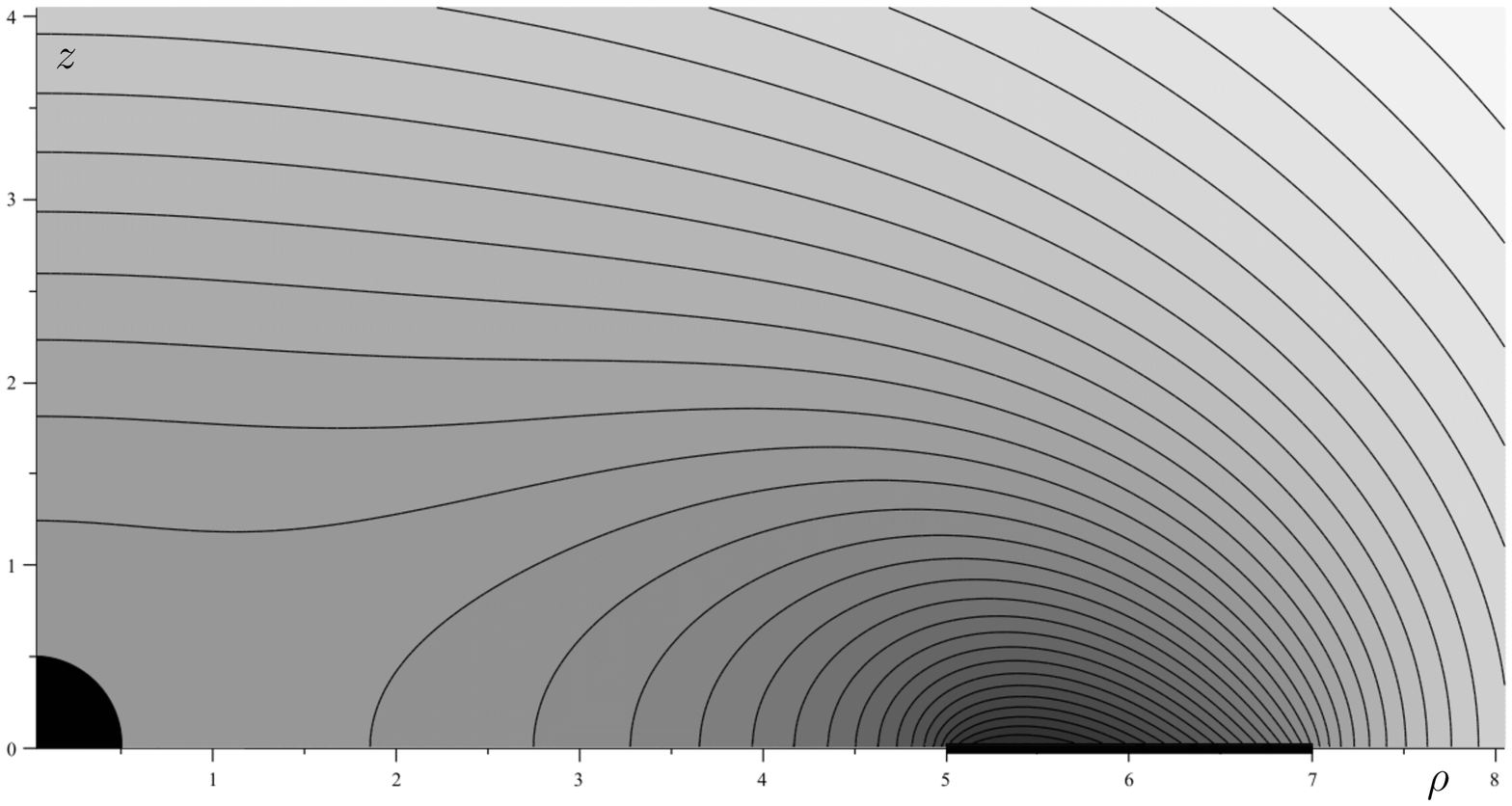}
\caption
{Meridional-plane contours of the dragging angular velocity $\omega$, as entirely given by the first-order perturbation due to the disc ($\omega_1$). The disc again stretches from $\rho\!=\!5M$ to $\rho\!=\!7M$ (as indicated by the thick black line). The angular velocity is everywhere positive, with light/dark shading indicating smaller/larger values (they reach about $0.0268/M$ at the disc and fall off to some $0.0047/M$ at top right of the plot); the black quarter-disc at $\rho^2+z^2\leq M^2/4$ represents the black hole. Both axes are given in the units of $M$. Since $\omega\!=\!\omega_1$ is proportional to $W$, the isolines have the same shape for any $W$, only their values scale with this rotational parameter.}
\label{fig:dragging}
\end{center}
\end{figure}

Differentiating the potential (\ref{nu1}), (\ref{LassBlitzer-potential}) across the disc plane and regarding that
$\nu_{,\theta}(\theta\!=\!\pi/2^-)=-r\nu_{,z}(z\!=\!0^+)$,
we find
\begin{equation}  \label{eq:sigma+P,nuz}
  \left.\frac{\partial V}{\partial\theta}\right|_{\theta={\frac{\pi}{2}}^-}\!\!
    = -2\pi MSx\,H(x'-x)
  \qquad \Longrightarrow \qquad
  \nu_{1,z}(z\!=\!0^+)
    = 2\pi S\left(1+\frac{M^2}{4r^2}\right)
    = \frac{2\pi SM}{r}\,\frac{1+v_0^2}{v_0^2} \;,
\end{equation}  
(the result only applies where the disc actually lies, i.e. at $x_{\rm in}\leq\frac{r}{M}\!+\!\frac{M}{4r}\leq x_{\rm out}$, elsewhere it is zero of course), hence, according to (\ref{nuz,jump,1}),
\begin{equation}  \label{sigmas-sum}
  \sigma_+\!+\sigma_-
    = \frac{\nu_{1,z}(z\!=\!0^+)}{2\pi}\,\frac{1-v_0^2}{1+v_0^2}
    = \frac{S}{4r^2}\,(4r^2\!-\!8Mr\!+\!M^2) \;.
\end{equation}

More involved is to find the equatorial limit of the normal gradient of $\omega_1$ (\ref{LassBlitzer-omega_1}). One finds that solely the term $Q_7 I_7$ contributes eventually, because $Q_5 I_5$ is zero from the beginning, the first separate term of (\ref{LassBlitzer-omega_1}) as well as $Q_0 I_0$ are independent of $\theta$, $Q_6 I_6$ is independent of $x'$ (and thus its ``out'' and ``in'' values subtract to zero), the $Q_2 I_2$ and $Q_3 I_3$ terms vanish in the $\theta\rightarrow(\pi/2)^-$ limit, and contributions of the terms $Q_1 I_1$ and $Q_4 I_4$ exactly cancel out each other in this limit. From $(Q_7 I_7)_{,\theta}$ one specifically obtains, in the above limit, non-zero terms
\[\frac{2Q_7\,(a_1+1)}{\sqrt{(a_1-1)^3(a_2+1)}}
  \left[-\Pi(n,k)\,\cos\theta
        +(1\!-\!\sin\theta)\,\frac{\partial\Pi(n,k)}{\partial\theta}\right],
  \qquad {\rm where} \qquad
  n=n(x,x',\theta):=\frac{2\sin\theta\,(a_1-1)}{(1+\sin\theta)(a_1-\sin\theta)}\]
and where the derivative of $\Pi(n,k)$ contributes solely through the term
$\frac{\Pi(n,k)}{2(1-n)}\,\frac{\partial n(x,x',\theta)}{\partial\theta}\,$.
Using the asymptotics
\[\Pi(n,k)\propto \frac{\pi\,\sqrt{n}}{2\,\sqrt{n-k^2}\,\sqrt{1-n}}+O(1)
  \qquad {\rm valid~for} \quad n\rightarrow 1^-\]
in both terms and making the equatorial limit, one finally finds a simple result
\begin{equation}  \label{eq:omegaz}
  \lim_{\theta\rightarrow\frac{\pi}{2}^-}
  \frac{\partial\omega_1}{\partial\theta}
  \left(=-r\lim_{z\rightarrow 0^+}\frac{\partial\omega_1}{\partial z}\right)
  =\frac{W}{2}\,\frac{x-1}{(x+1)^3}
  =8W M^2 r^2\;\frac{(2r-M)^2}{(2r+M)^6} \;,
\end{equation}
which, when compared against
\[\lim_{z\rightarrow 0^+}\frac{\partial\omega_1}{\partial z}
    = -8\pi(\sigma_+-\sigma_-)\,\frac{\Omega_0}{1-v_0^2}
    = -8\pi(\sigma_+-\sigma_-)\,
       \frac{8r\,\sqrt{Mr}}{(2r+M)^3}\,\frac{(2r-M)^2}{4r^2-8Mr+M^2}\]
given by (\ref{omegaz,jump,1}), implies that
\begin{equation}  \label{sigmas-diff}
  \sigma_+-\sigma_- = \frac{WM^2}{8\pi\,\sqrt{Mr}}\,\frac{4r^2-8Mr+M^2}{(2r+M)^3} \;.
\end{equation}
By adding (\ref{sigmas-sum}) and (\ref{sigmas-diff}), we have then
\begin{equation}
  \sigma_{\pm}=\frac{4r^2-8Mr+M^2}{8r^2}
               \left[S \pm \frac{W\,(Mr)^{3/2}}{2\pi\,(2r+M)^3}\right].
\end{equation}

To find the one-stream disc characteristics, one first combines (\ref{nuz,jump,1}) and (\ref{zetaz,jump,1}), which yields
\begin{equation}  \label{v^2}
  v^2 = \frac{\sigma v_0^2-P}{\sigma-Pv_0^2}
      = \frac{4Mr\sigma-(2r-M)^2 P}{(2r-M)^2\sigma-4MrP} \;.
\end{equation}
Substituting this for $v^2$ in (\ref{omegaz,jump,1}) leads to
\[(\sigma v_0^2-P)(\sigma-Pv_0^2)=(\sigma_+-\sigma_-)^2\,v_0^2\]
and from there, using $\sigma\!-\!P=\sigma_+\!+\!\sigma_-$, we find
\begin{align}
  \sigma & = +\frac{\sigma_+\!+\!\sigma_-}{2} +
              \sqrt{\left(\frac{\sigma_+\!+\!\sigma_-}{2}\right)^2
                    +\frac{4\,\sigma_+\sigma_- v_0^2}{(1\!-\!v_0^2)^2}}
           = +\frac{\sigma_+\!+\!\sigma_-}{2} +
              \sqrt{\left(\frac{\sigma_+\!+\!\sigma_-}{2}\right)^2
                    +\frac{16Mr\,(2r\!-\!M)^2\,\sigma_+\sigma_-}{(4r^2\!-\!8Mr\!+\!M^2)^2}} \;, \\
  P      & = -\frac{\sigma_+\!+\!\sigma_-}{2} +
              \sqrt{\left(\frac{\sigma_+\!+\!\sigma_-}{2}\right)^2
                    +\frac{4\,\sigma_+\sigma_- v_0^2}{(1\!-\!v_0^2)^2}}
           = -\frac{\sigma_+\!+\!\sigma_-}{2} +
              \sqrt{\left(\frac{\sigma_+\!+\!\sigma_-}{2}\right)^2
                    +\frac{16Mr\,(2r\!-\!M)^2\,\sigma_+\sigma_-}{(4r^2\!-\!8Mr\!+\!M^2)^2}} \;.
\end{align}
Note that when substituting these $\sigma$ and $P$ back to (\ref{v^2}), the resulting $v$ given by square root of the latter is to be taken with $+/-$ sign in case that $\sigma_+\!>\!\sigma_-\,$/$\,\sigma_+\!<\!\sigma_-\,$.

In order to demonstrate that the procedure really works and, in particular, to illustrate the role of the parameters $S$ and $W$, let us consider a disc spanning between the radii $\rho_-\!=\!5M$ and $\rho_+\!=\!7M$, surrounding a black hole of mass $M$, and let us plot the results for several different values of $S$ and/or $W$. Note that both $S$ and $W$ have the dimension of 1/length, practically $1/M$. Note also that in order to emphasize their effect, we ignore here the assumption that the perturbation should be very small. More precisely, the first-order perturbation of the gravitational potential is in fact {\em not} restricted by this assumption, because rotation/dragging only enters in the second order, so the change of $\nu$ can be understood as an {\em exact} superposition {\em within the non-rotating, static class}. The potential $\nu$ (the sum of the Schwarzschild background $\nu_0$ and the perturbation $\nu_1$) is shown in figure \ref{fig:potential}; it behaves in an expected manner, namely the disc effect grows with increasing $S$ (and it is independent of $W$). The dragging angular velocity amounts to $\omega\!=\!\omega_1$ (since $\omega_0\!=\!0$), so the level-contour shape is in fact fixed and does not change with parameters (only the level {\em values} do change, in particular they scale with $W$); this is shown in figure \ref{fig:dragging}.

\begin{figure}[ht]
\includegraphics[width=\textwidth]{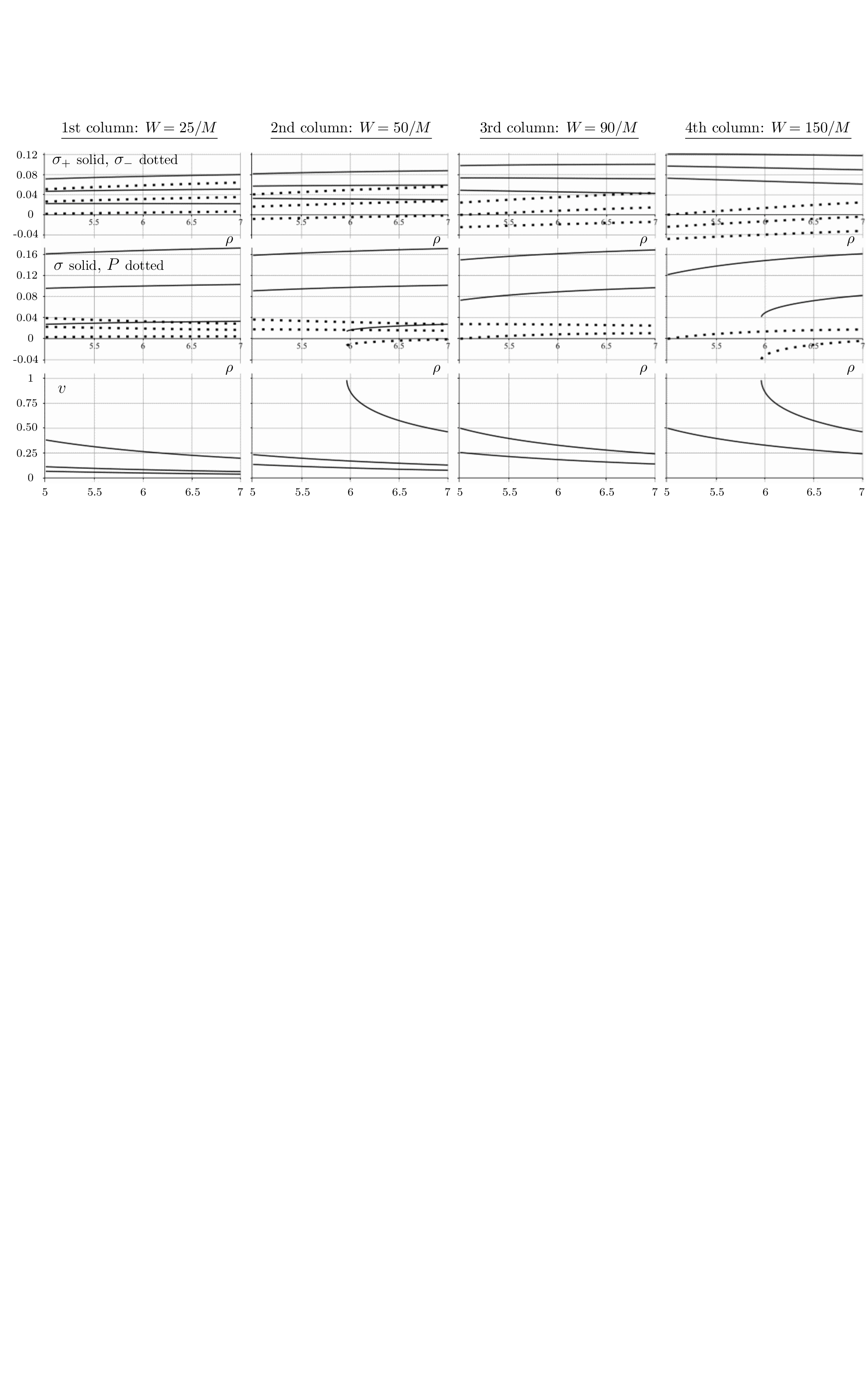}
\caption
{Parameters of the disc's double-stream and single-stream interpretations, plotted as functions of the Weyl radius for a disc stretching between $\rho\!=\!5M$ and $\rho\!=\!7M$ and for several combinations of the parameters $S$ (Newtonian density) and $W$ (scaling the disc rotation). The first \!/\! second \!/\! third \!/\! fourth columns represent the cases given by $W\!=\!25$ \!/\! $50$ \!/\! $90$ \!/\! $150$, with the first row showing densities $\sigma_+$ (solid line), $\sigma_-$ (dotted line) of the double-stream, counter-rotating dust interpretation, the second row showing density $\sigma$ (solid line) and azimuthal pressure $P$ (dotted line) of the single-stream fluid interpretation, and the third row showing the corresponding ``bulk'' velocity of the fluid $v$ (solid). For each of the values of $W$, three different values of $S$ have been chosen, corresponding to discs of different mass -- $S\!=\!0.04$, $0.12$ and $0.2$. The density and pressure curves obtained for higher $S$ are higher (given by larger values), whereas the corresponding bulk velocities decrease with growing $S$. The ranges of the chosen $W$ and $S$ are largely out of the scope of the linear perturbation, but this is in order to illustrate clearly what they represent. Since all the parameters (densities, pressure, velocity) should be real and positive, and the velocity $v$ has to be $<\!1$ in addition, one sees immediately that i) the discs with a given mass cannot bear however large angular momentum (with growing $W$, the density $\sigma_-$ of the double-stream interpretation -- and consequently also the single-stream parameters -- tend to be negative); and, ii) since large $W$ implies the need for high orbital velocities, for too large $W$ the orbital interpretation would have to involve superluminal motion (in the bottom row, from left to right, one sees that all three \!/\! two \!/\! two \!/\! one of the shown cases are ``physical'' in this respect). The densities $S$, $W$, $\sigma_+$, $\sigma_-$, $\sigma$ as well as the pressure $P$ have the dimension of 1/length and their values are in the units of $1/M$, the speed $v$ is dimensionless.}
\label{fig:disc-params}
\end{figure}

Now for parameters of the double-stream and single-stream interpretations, i.e., $\sigma_\pm$ and $v_\pm$, and $\sigma$, $P$ and $v$, respectively. In fact the orbital velocities $v_\pm$ of the double-stream picture need not be computed, namely, in the first perturbation order they are the same as in a pure Schwarzschild field, because the field equations (\ref{eq:EFE2}) and (\ref{eq:EFE3}) have right-hand sides proportional to the surface densities which are themselves of linear order. The other quantities are plotted in figure \ref{fig:disc-params} for several choices of the $S$ and $W$ parameters. One sees there that an increasing value of $W$ makes $\sigma_+$ and $\sigma_-$ more and more different, with $\sigma_-$ finally becoming negative, which marks limits of the (counter-)rotating interpretation. Physically, such a situation means that, for a given mass, the disc has too much angular momentum. Naturally, this is also accompanied by a need for too high orbital velocity; such a circumstance can be ``remedied'' by increasing the mass, i.e. $S$.

\subsection{Mass and angular momentum of the disc}
\label{section:M,J}

The total mass and angular momentum of a stationary and axially symmetric space-time can be found from Komar integrals, given by the Killing vector fields $\eta^\mu\!=\!\frac{\partial x^\mu}{\partial t}$ and $\xi^\mu\!=\!\frac{\partial x^\mu}{\partial \phi}$. In our case (involving two sources) the integrals over spatial infinity can be split into contributions from the black hole $M$ and $J$ (given by integration over the horizon H) and from the disc (integrated over some space-like hypersurface $\Sigma$ covering the black-hole exterior),
\begin{align}
  {\rm mass}
   &= \frac{1}{8\pi}\lim_{S\rightarrow\infty}\oint\limits_S
      \eta^{\mu;\nu}{\rm d}S_{\mu\nu}
    = \frac{1}{8\pi}\oint\limits_{\rm H}\eta^{\mu;\nu}{\rm d}S_{\mu\nu}
      -\int\limits_{\Sigma>{\rm H}}(2T^\mu_\nu\eta^\nu\!-T\eta^\mu)\,{\rm d}\Sigma_\mu
    = M + \int\limits_{\Sigma>{\rm H}}(T^i_i-T^t_t)\sqrt{-g}\;{\rm d}^3x \,, 
   \label{total-mass} \\
  {\rm ang.m.}
   &= -\frac{1}{16\pi}\lim_{S\rightarrow\infty}\oint\limits_{S}
       \xi^{\mu;\nu}{\rm d}S_{\mu\nu}
    = -\frac{1}{16\pi}\oint\limits_{\rm H}\xi^{\mu;\nu}{\rm d}S_{\mu\nu}
      +\frac{1}{2}\int\limits_{\Sigma>{\rm H}}(2T^\mu_\nu\xi^\nu\!-T\xi^\mu)\,{\rm d}\Sigma_\mu
    = J + \int\limits_{\Sigma>{\rm H}} T^t_\phi\,\sqrt{-g}\;{\rm d}^3x \,,
   \label{total-angular-momentum}
\end{align}
where we have finally employed the ``Killing'' coordinates $t$ and $\phi$ in which $\eta^\mu=\delta^\mu_t$, $\xi^\mu=\delta^\mu_\phi$, and a natural space-like section $\Sigma=\{t\!=\!{\rm konst}\}$ (which corresponds to ${\rm d}\Sigma_\mu=\delta^t_\mu\sqrt{-g}\,{\rm d}^3x$). Clearly $(T^i_i\!-\!T^t_t)$ -- or actually $e^{2\lambda-2\nu}(T^i_i\!-\!T^t_t)$ -- plays the same role as mass density in the Newtonian theory. Note also that the second part of the mass integral represents what is sometimes called Tolman's formula.

Choosing $B\!=\!1$, the metric determinant reads $-g=\rho^2 e^{4\lambda-4\nu}$, so, if considering the thin $\{z\!=\!0\}\,$-disc as source and thus having $T^\mu_\nu\sqrt{-g}=S^\mu_\nu\rho\,\delta(z)$, the disc mass and angular momentum (denoted by ${\cal M}$ and ${\cal J}$) come out as
\[{\cal M}=2\pi\int\limits_{\rm disc}(S^\phi_\phi-S^t_t)\,\rho\,
           {\rm d}\rho\,,
  \qquad
  {\cal J}=2\pi\int\limits_{\rm disc}S^t_\phi\,\rho\,{\rm d}\rho\,.\]
Substituting here for (\ref{S-components}) with $u^\phi=u^t\Omega$ and $u_\phi=g_{\phi\phi}u^t(\Omega\!-\!\omega)$, we have
\begin{equation}
  S^t_\phi = (\sigma+P)(u^t)^2 g_{\phi\phi}(\Omega-\omega), \qquad
  S^\phi_\phi-S^t_t
      = (\sigma+P)(u^t)^2\left[e^{2\nu}+g_{\phi\phi}(\Omega^2-\omega^2)\right]
      = \sigma+P+2\Omega S^t_\phi\,,
\end{equation}
hence
\begin{align}
  {\cal M} &= 2\pi\int\limits_{\rm disc}
              (\sigma+P+2\Omega S^t_\phi)\,\rho\,{\rm d}\rho \;, \\ 
  {\cal J} &= 2\pi\int\limits_{\rm disc}(\sigma+P)\,
              \frac{\rho e^{-2\nu}(\Omega-\omega)}{1-\rho^2 e^{-4\nu}(\Omega-\omega)^2}\;
              g_{\phi\phi}\,{\rm d}\rho\;;
\end{align}
in the special case of $\Omega\!=\!{\rm const}$ the first integral amounts to
\[{\cal M}=2\pi\int\limits_{\rm disc}(\sigma+P)\,\rho\,{\rm d}\rho
           +2\Omega{\cal J}\,.\]

One can alternatively express, in the integrals, $S^\mu_\nu$ in terms of jumps of the normal fields, according to (\ref{nuz,jump})--(\ref{omegaz,jump}). Setting $B\!=\!1$ again, we thus have
\[S^t_\phi=-\frac{1}{8\pi}\,\rho^2 e^{-4\nu}\omega_{,z} \;, \qquad
  S^\phi_\phi-S^t_t=\frac{\nu_{,z}}{2\pi}+2\omega S^t_\phi \;,\]
hence
\begin{equation}
  {\cal M}= \int\limits_{\rm disc}
            \left(\nu_{,z}-\frac{1}{2}\,\rho^2 e^{-4\nu}\omega\omega_{,z}\right)
            \rho\,{\rm d}\rho \;,
  \qquad\quad
  {\cal J}= -\frac{1}{4}\int\limits_{\rm disc}
             \rho^3 e^{-4\nu}\omega_{,z}\,{\rm d}\rho
\end{equation}
(where $\omega_{,z}$ and $\nu_{,z}$ are evaluated at $z\rightarrow 0^+$).
Since the second term in ${\cal M}$ is $O(\lambda^2)$, in the linear order one is left with just
\begin{equation}
  {\cal M}_1 = \int\limits_{\rm disc}\nu_{1,z}(z\!\rightarrow\!0^+)\,\rho\;{\rm d}\rho \,,
  \qquad\qquad
  {\cal J}_1 = -\frac{1}{4}\int\limits_{\rm disc}
               \rho^3 e^{-4\nu_0}\,\omega_{1,z}(z\!\rightarrow\!0^+)\;{\rm d}\rho\;.
\end{equation}
Finally, one has to realize that the above formulas hold for $B\!=\!1$, whereas our results for $\nu_1$ and $\omega_1$ have been derived with $B=1-\frac{M^2}{4r^2}\,$. However, adapting the integrals to the latter choice practically means just to write $B\rho$ instead of $\rho$ in both integrands, reaching
\begin{equation}
  {\cal M}_1 = \int\limits_{\rm disc}B\rho\,\nu_{1,z}(z\!\rightarrow\!0^+)\;{\rm d}\rho \,,
  \qquad\qquad
  {\cal J}_1 = -\frac{1}{4}\int\limits_{\rm disc}
               B^3\rho^3 e^{-4\nu_0}\,\omega_{1,z}(z\!\rightarrow\!0^+)\;{\rm d}\rho\;.
\end{equation}

Considering specifically the above constant-density disc, it is clear from (\ref{LassBlitzer-potential}) and (\ref{LassBlitzer-omega_1}) that ${\cal M}_1$ is scaled by the free parameter $S$ while ${\cal J}_1$ is scaled by the second parameter $W$. We have from (\ref{eq:sigma+P,nuz}) and (\ref{eq:omegaz}), at $z\!\rightarrow\!0^+$ (i.e., $\theta\!\rightarrow\pi/2^-$),
\begin{equation}
  \nu_{1,z}\rho = -\nu_{1,\theta}
                = \pi S\left(2r+\frac{M^2}{2r}\right) \,,
  \qquad
  \omega_{1,z}\rho = -\omega_{1,\theta}
                   = -8WM^2 r^2\,\frac{(2r-M)^2}{(2r+M)^6} \;,
\end{equation}
which yields, after using $\rho(\theta\!=\!\pi/2)=r$ and
$e^{-4\nu_0}=\frac{(2r+M)^4}{(2r-M)^4}\,$,
\begin{align}
  {\cal M}_1 
     &= \pi S \int\limits_{r_{\rm in}}^{r_{\rm out}}
              \left(2r+\frac{M^2}{2r}\right)\left(1-\frac{M^2}{4r^2}\right)
              {\rm d}r
      = \pi S \left[r_{\rm out}^2-r_{\rm in}^2
                    -\frac{M^4}{16}\left(\frac{1}{r_{\rm in}^2}-\frac{1}{r_{\rm out}^2}\right)\right], \\
  {\cal J}_1
     &= \frac{WM^2}{8} \int\limits_{r_{\rm in}}^{r_{\rm out}}\left(1-\frac{M^2}{4r^2}\right){\rm d}r
      = \frac{WM^2}{8} \left[r_{\rm out}-r_{\rm in}-
                             \frac{M^2}{4}\left(\frac{1}{r_{\rm in}}-\frac{1}{r_{\rm out}}\right)\right]. 
\end{align}
The same results follow from the asymptotics (\ref{nu1,r=infty}) and (\ref{omega1,r=infty}), if regarding the general behaviour (in an asymptotically flat space-time)
\[\nu(r\!\rightarrow\infty) \propto -\frac{M_1+{\cal M}_1}{r} \;,
  \qquad\qquad
  \omega(r\!\rightarrow\infty) \propto 2\,\frac{J_1+{\cal J}_1}{r^3} \;.\]

The last statement implies that i) the black-hole mass remains $M$ (which should be so in the linear perturbation order), and that ii) the whole angular momentum of the system (inferred from the asymptotic behaviour of $\omega$) is being carried by the disc, hence that the angular momentum of the hole remains zero, $J\!=\!0$. Actually, we can verify this directly by computing the respective Komar integral over the horizon. It is known that the latter can be rewritten (see e.g. \citealt{bib:Will1974}, equation (21)) as
\[J = -\frac{k^4}{2^7}
       \int\limits_0^\pi 
       \left[B^3\omega_{,r}\,e^{-4\nu}\right]_{r=k/2}\,\sin^3\theta\;{\rm d}\theta \,,\]
where in our first-order case one takes $\nu\!=\!\nu_0$, $\omega\!=\!\omega_1$ and $k\!=\!M$.
Inspecting the radial gradient $\omega_{1,r}$ of (\ref{LassBlitzer-omega_1}), one finds that all the terms of this expression {\em individually} vanish at the horizon (irrespectively of $\theta$), so the above formula really yields $J\!=\!0$. Hence, the perturbed black hole is rotating with respect to the asymtotic inertial frame with the non-zero (positive) angular velocity (\ref{omega1,H}), but has a zero angular momentum, which means that it is just being ``carried along'' by dragging primarily induced by the disc. We stress that this feature is {\em not} a necessary outcome of the perturbation procedure, namely it is a consequence of our choice of the constants $J_l$ (namely $J_l\!=\!0$) which can be employed to fix the black-hole spin in the solution (\ref{eq:EFE3_DecInhomogSol}) of the inhomogeneous perturbation equation (\ref{eq:EFE3_Dec}); see the discussion in section \ref{sec:WillComp}.

\section{Concluding remarks}
	
We have shown that the procedure suggested by \cite{bib:Will1974}, originally employed to determine the gravitational perturbation of a Schwarzschild black hole by a slowly rotating and light thin ring, can be also applied to the perturbation due to a thin disc. However, concerning the bad numerical properties of the series involved, we have dropped the angular expansion and rather expressed the Green functions (perturbations due to a thin ring) in a closed form. Such expressions bring more complex special functions, but these can be evaluated effectively with rapidly converging algorithms, so the resulting numerical convergence is much better.

Using the proposed closed-form Green functions, one can in principle obtain an arbitrary order of the perturbation. However, a numerical treatment is necessary in order to analyse specific results, as illustrated on a simple example of the ``uniform-density'' disc in section \ref{ch:Disc}.

An important point has been to show that the series involved in computation of the Green functions converge (section \ref{ch:Convergence} and the accompanying paper by \cite{bib:math-part}). However, what still remains to be answered is whether the perturbation scheme is effective to {\em any} order, namely whether the perturbation expansion (\ref{eq:Expansion}) with parameter proportional to the external-source mass converges. Regarding the structure of the Green functions and the speed of their convergence to zero at infinity, as well as the structure of source terms of higher perturbation orders, one conjectures that the expansion converges at least for some positive disc masses.

Various properties of the obtained solution could be studied now, for instance, deformation of the geometry (as represented by curvature invariants, in particular), deformation of the (originally spherical) horizon, perturbation of the properties of stationary circular motion or influence on geodesic structure. In particular, it will be interesting to see how the perturbation influences the location of important circular equatorial geodesics (mainly of the innermost stable one, usually abbreviated as ISCO), because this should indicate how the actual quasi-stationary accretion disc may differ from its test-matter model.
Also, \`a propos, one could consider a different type of disc (a different density distribution) -- preferably close to what follows from models of accretion onto astrophysical black holes -- and compare the results with what has been found here for the simplest case of constant density.

\subsection{Comparison with black-hole--disc configurations found numerically}

Another obvious option is to compare the perturbative solution with the results of numerical treatment of similar source configurations. The most similar of these -- a hole with a thin finite annular equatorial disc -- was studied by \cite{bib:Lanza1992}, while \cite{bib:NishidaE1994} considered a hole with a thick toroid. More recently, \cite{bib:AnsorgP2005} used a different numerical method to compute stationary and axisymmetric configurations of uniformly rotating constant-density toroids around black holes. They specifically used these solutions to demonstrate that both the central hole and the surrounding toroid may in some cases have {\em negative} Komar masses \citep{bib:AnsorgP2006}. Yet another codes for studying self-gravitating matter around black holes have been developed, and specifically used to find stationary thick-toroid configurations, by \cite{bib:Shibata2007}, \cite{bib:MonteroFS2008} and \cite{bib:Stergioulas2011}. Finally, \cite{bib:KarkowskiMMPX2016} analyzed such rotating black-hole--toroid systems in the first post-Newtonian approximation.

The possibility to compare our results with the above ``exact'' numerical configurations (in particular the thin discs by Lanza) is very limited, mainly because our first-order perturbation only represents gravitation of the disc, not its {\em self}-gravitation (no back effect of the source on itself through its field); more generally speaking, the solution does not incorporate any non-linearity of the Einstein equations.
There are also other, more definite differences. The numerical results of \cite{bib:Lanza1992} were obtained by numerical ``relaxation'' of initial configurations provided by ``squeezing'' the well known (analytical) {\em test} thick-disc models in a given Kerr background. More specifically, assuming constant ratio of angular momentum to energy (with respect to infinity) throughout the disc, in our notation
\[{\rm const} = \frac{u_\phi}{-u_t}
              = -\frac{g_{t\phi}+g_{\phi\phi}\Omega}{g_{tt}+g_{t\phi}\Omega}
              = \left(\omega+\frac{e^{2\nu}}{B\rho\,v}\right)^{\!-1} , \]
and choosing the inner disc radius and the proportionality constant appearing in the polytropic equation of the disc-gas state, the initial surface density and outer radius of the disc are obtained. Generally, with the increase of that constant, the surface density (as well as pressure) and mass of the resulting disc rise quickly, while the outer disc radius grows with the specific angular momentum of the disc matter.
In contrast to Lanza's constraint of constant specific angular momentum, and the disc's surface density and outer radius derived accordingly, we have exemplified our perturbation procedure on a disc with prescribed and constant Newtonian surface density (while non-constant angular momentum) and with both inner and outer radii prescribed as well.

Lanza illustrated his numerical scheme on two groups of configurations sequences, one containing a rapidly rotating hole (specified by its horizon area and angular momentum) and one with a slowly rotating hole (in this case specified by its horizon angular velocity $\omega_{\rm H}$ and isotropic radius $k/2$). Focusing naturally only to the latter, one finds solution sequences for three different rotation choices in the cited paper: $\omega_{\rm H}\!=\!0.00025/M$, $\omega_{\rm H}\!=\!0.0025/M$ and $\omega_{\rm H}\!=\!0.025/M$, in all cases with $k\!=\!M$ and $r_{\rm in}\!=\!8M$; each sequence is characterized by a certain fixed value of the angular momentum to energy ratio (constant throughout the disc) and was generated by gradual increase of the disc mass. With the increasing disc mass, also the total mass of the system were found to increase (almost linearly), while the total angular momentum and the horizon area $A_{\rm H}$ were increasing slightly faster, with the horizon surface gravity decreasing consequently according to the generic relation $\kappa_{\rm H}=4\pi k/A_{\rm H}$. One special point Lanza examined on slowly rotating configurations was that the black-hole rotational angular momentum may decrease to negative values when the disc angular momentum is being increased (while $\omega_{\rm H}$ is kept fixed). This happens when the disc is ``overtaking'' the black hole in the sense that the combined rotational-dragging effect is stronger than the effect due to the hole alone (such a circumstance was already pointed out by \citealt{bib:Will1974}). Let us remind that in our case the black hole was effectively {\em set} to keep zero angular momentum (while acquiring non-zero angular velocity) in the perturbation, as confirmed at the end of section \ref{section:M,J}.

Regarding the differences between assumptions of the above numerical treatment and our perturbative one, the comparison of results obtained by these two methods is going to be rather problematic. Anyway, we now plan to compute various more specific parameters of the disc considered as an example in section \ref{ch:Disc} (and the following ones), in order to possibly return to this point. Let us thus conclude with one particular surprising observation made by \cite{bib:Lanza1992} which should be {\em simple} to compare: for slowly rotating black holes, Lanza found that the presence of the disc can make the horizon's polar proper circumference {\em larger} than the equatorial one, which would suggest that the horizon becomes prolate (along the symmetry axis). This goes against common experience that the black holes stretch towards external sources of gravity (cf. \citealt{bib:NishidaE1994} and \citealt{bib:AnsorgP2005} who always got oblate horizons).

\acknowledgments
We are grateful to P. Kotla\v{r}\'{\i}k for reading the paper and for checking most of the formulas, and also to T. Ledvinka for interest and useful comments.
O.S. thanks for support from the grant GACR-17/13525S of the Czech Science Foundation.


\begin{thebibliography}{}

\bibitem[Abramowicz et. al(1984)]{bib:Abr1984}
   Abramowicz M. A., Curir A., Schwarzenberg-Czerny A., \& Wilson R. E., 1984,
   Self-gravity and the global structure of accretion discs,
   MNRAS 208, 279
\bibitem[Ansorg \& Petroff(2005)]{bib:AnsorgP2005}
   Ansorg M. \& Petroff D., 2005,
   Black holes surrounded by uniformly rotating rings,
   Phys. Rev. D 72, 024019
\bibitem[Ansorg \& Petroff(2006)]{bib:AnsorgP2006}
   Ansorg M. \& Petroff D., 2006,
   Negative Komar mass of single objects in regular, asymptotically flat spacetimes,
   Class. Quantum Grav. 23, L81
\bibitem[Baranov(2006)]{bib:Baranov2006}
   Baranov A. S., 2006,
   On series containing products of Legendre polynomials,
   Math. Notes 80, 167
\bibitem[Bardeen(1973)]{bib:Bardeen1973}
   Bardeen J. M., 1973,
   Rapidly Rotating Stars, Disks, and Black Holes,
   in DeWitt C., DeWitt B. S., eds,
   Black Holes (Les Houches 1972), New York, Gordon and Breach, p. 241
\bibitem[Bi\v{c}\'ak \& Ledvinka(1993)]{bib:BicLed1993}
   Bi\v{c}\'ak J. \& Ledvinka T., 1993,
   Relativistic disks as sources of the Kerr metric,
   Phys. Rev. Lett. 71, 1669
\bibitem[Bi\v{c}\'ak et al.(1993)]{bib:Bic1993}
   Bi\v{c}\'ak J., Lynden-Bell D., \& Katz J., 1993,
   Relativistic disks as sources of static vacuum spacetimes,
   Phys. Rev. D 47, 4334
\bibitem[Bret\'on et. al(1997)]{bib:Breton1997}
   Bret\'on N., Denisova T. E., \& Manko V. S., 1997,
   A Kerr black hole in the external gravitational field,
   Phys. Lett. A 230, 7
\bibitem[Bret\'on et. al(1998)]{bib:Breton1998}
   Bret\'on N., Garc\'{\i}a A. A., \& Manko V. S., 1998,
   Arbitrarily deformed Kerr-Newman black hole in an external gravitational field,
   Phys. Rev. D 57, 3382
\bibitem[Chaudhuri \& Das(1997)]{bib:ChauDas1997}
   Chaudhuri S. \& Das K. C., 1997,
   Axially symmetric metrics from Laplace's seed by inverse scattering method,
   J. Math. Phys. 38, 5792
\bibitem[Chrzanowski(1976)]{bib:Chrzanowski1976}
   Chrzanowski P. L., 1976,
   Applications of metric perturbations of a rotating black hole -- Distortion of the event horizon,
   Phys. Rev. D 13, 806
\bibitem[\v{C}\'{\i}\v{z}ek \& Semer\'ak(2009)]{bib:CizSem2009}
   \v{C}\'{\i}\v{z}ek P. \& Semer\'ak O., 2009,
   Thin-disc Perturbation of a Schwarzschild Black Hole,
   in \v{S}afr\'ankov\'a J., Pavl\r{u} J., eds,
   WDS 2009, Proc. 18th Annual Conf. of Doctoral Students,
   Matfyzpress, Prague, p. 38
\bibitem[\v{C}\'{\i}\v{z}ek(2011)]{bib:Ciz2011}
   \v{C}\'{\i}\v{z}ek P., 2011,
   Linear perturbations of a Schwarzschild black hole by thin disc,
   J. Phys. Conf. Ser. 314, 012071
\bibitem[\v{C}\'{\i}\v{z}ek(in preparation)]{bib:math-part}
   \v{C}\'{\i}\v{z}ek P.,
   Perturbation of a Schwarzschild black hole due to a rotating thin disc: Mathematical details,
   in preparation
\bibitem[Demianski(1976)]{bib:Demianski1976}
   Demianski M., 1976,
   Stationary axially symmetric perturbations of a rotating black hole,
   Gen. Rel. Grav. 7, 551
\bibitem[Garc\'{\i}a-Reyes \& Gonz\'alez(2004)]{bib:Garcia2004}
   Garc\'{\i}a-Reyes G. \& Gonz\'alez G. A., 2004,
   Counterrotating perfect fluid discs as sources of electrovacuum static spacetimes,
   Class. Quantum Grav. 21, 4845
\bibitem[Gonz\'alez \& Espitia(2003)]{bib:GonzEspi2003}
   Gonz\'alez G. A. \& Espitia O. A., 2003,
   Relativistic static thin disks: The counterrotating model,
   Phys. Rev. D 68, 104028
\bibitem[Grandcl\'ement \& Novak(2008)]{bib:GrandNovak2008}
   Grandcl\'ement P. \& Novak J., 2008,
   Spectral Methods for Numerical Relativity,
   Living Rev. Relativity 12, 1
\bibitem[Hod(2014)]{bib:Hod2014}
   Hod S., 2014,
   Self-gravitating ring of matter in orbit around a black hole: the innermost stable circular orbit,
   Eur. Phys. J. C 74, 2840
\bibitem[Hod(2015)]{bib:Hod2015}
   Hod S., 2015,
   Dragging of inertial frames in the composed black-hole--ring system,
   Eur. Phys. J. C 75, 541
\bibitem[Karkowski et al.(2016)]{bib:KarkowskiMMPX2016}
   Karkowski J., Mach P., Malec E., Pir\'og M., \& Xie N., 2016,
   Rotating systems, universal features in dragging and antidragging effects, and bounds of angular momentum,
   Phys. Rev. D 94, 124041
\bibitem[Kato et. al(2008)]{bib:Kato2008}
   Kato S., Fukue J., \& Mineshige S., 2008,
   Black-Hole Accretion Disks. Towards a New Paradigm,
   Kyoto Univ. Press, Kyoto
\bibitem[Klein \& Richter(1999)]{bib:KleinRich1999}
   Klein C. \& Richter O., 1999,
   Exact relativistic gravitational field of a stationary counterrotating dust disk,
   Phys. Rev. Lett. 83, 2884
\bibitem[Klein \& Richter(2005)]{bib:KleinRich2005}
   Klein C. \& Richter O., 2005,
   Ernst Equation and Riemann Surfaces: Analytical and Numerical Methods, Lect. Notes Phys. 685,
   Springer, Berlin
\bibitem[Krori \& Bhattacharjee(1990)]{bib:KroriBat1990}
   Krori K. D. \& Bhattacharjee R., 1990,
   A Kerr object embedded in a gravitational field. II,
   J. Math. Phys. 31, 147
\bibitem[Lamberti \& Hamity(1989)]{bib:LambHam1989}
   Lamberti P. W. \& Hamity V. H., 1989,
   Counterrotating disks of collisionless particles and the hoop conjecture,
   Gen. Rel. Grav. 21, 869
\bibitem[Lanza(1992)]{bib:Lanza1992}
   Lanza A., 1992,
   Self-gravitating thin disks around rapidly rotating black holes,
   ApJ 389, 141
\bibitem[Lass \& Blitzer(1983)]{bib:LassBlitzer1983}
   Lass H. \& Blitzer L., 1983,
   The gravitational potential due to uniform disks and rings,
   Celest. Mech. 30, 225
\bibitem[Lynden-Bell \& Pineault(1978)]{bib:LyndenBell1978}
   Lynden-Bell D. \& Pineault S., 1978,
   Relativistic disks -- 1. Counter rotating disks,
   MNRAS 185, 679
\bibitem[Meinel et al.(2008)]{bib:Meinel2008}
   Meinel R., Ansorg M., Kleinw\"{a}chter A., Neugebauer G., \& Petroff D., 2008,
   Relativistic Figures of Equilibrium,
   Cambridge Univ. Press, Cambridge
\bibitem[Montero et al.(2008)]{bib:MonteroFS2008}
   Montero P. J., Font J. A., \& Shibata M., 2008,
   Nada: A new code for studying self-gravitating tori around black holes,
   Phys. Rev. D 78, 064037
\bibitem[Morgan \& Morgan(1969)]{bib:MorganMorgan1969}
   Morgan T. \& Morgan L., 1969,
   The gravitational field of a disk,
   Phys. Rev. 183, 1097
\bibitem[Nishida \& Eriguchi(1994)]{bib:NishidaE1994}
   Nishida S. \& Eriguchi Y., 1994,
   A general relativistic toroid around a black hole,
   ApJ 427, 429
\bibitem[Olver et. al.(2010)]{bib:DLMF}
   Olver F. W. J. et. al., 2010,
   NIST Handbook of Mathematical  Functions,
   Cambridge Univ. Press, Cambridge
\bibitem[Ruggiero(2016)]{bib:Ruggiero2016}
   Ruggiero M. L., 2016,
   Gravitomagnetic field of rotating rings,
   Astrophys. Space. Sci. 361, 140
\bibitem[Sano \& Tagoshi(2014)]{bib:SanoTagoshi2014}
   Sano Y. \& Tagoshi H., 2014,
   Gravitational field of a Schwarzschild black hole and a rotating mass ring,
   Phys. Rev. D 90, 044043
\bibitem[Semer\'ak(2002)]{bib:Semer2002}
   Semer\'ak O., 2002,
   Thin disc around a rotating black hole, but with support in-between,
   Class. Quantum Grav. 19, 3829
\bibitem[Semer\'ak(2003)]{bib:Semer2003}
   Semer\'ak O., 2003,
   Gravitating discs around a Schwarzschild black hole: III,
   Class. Quantum Grav. 20, 1613
\bibitem[Semer\'ak(2004)]{bib:Semer2004}
   Semer\'ak O., 2004,
   Exact power-law discs around static black holes,
   Class. Quantum Grav. 21, 2203
\bibitem[Semer\'ak(2016)]{bib:Semer2016}
   Semer\'ak O., 2016,
   Static axisymmetric rings in general relativity: How diverse they are,
   Phys. Rev. D 94, 104021
\bibitem[Shibata(2007)]{bib:Shibata2007}
   Shibata M., 2007,
   Rotating black hole surrounded by self-gravitating torus in the puncture framework,
   Phys. Rev. D 76, 064035
\bibitem[Stergioulas(2011)]{bib:Stergioulas2011}
   Stergioulas N., 2011,
   An improved method for constructing models of self-gravitating tori around black holes,
   Int. J. Mod. Phys. D 20, 1251
\bibitem[Sukov\'a \& Semer\'ak(2013)]{bib:SukovaSem2013}
   Sukov\'a P. \& Semer\'ak O., 2013,
   Free motion around black holes with discs or rings: between integrability and chaos -- III,
   MNRAS 436, 978
\bibitem[Tomimatsu(1984)]{bib:Tomi1984}
   Tomimatsu A., 1984,
   Distorted rotating black holes,
   Phys. Lett. A 103, 374
\bibitem[Will(1974)]{bib:Will1974}
   Will C. M., 1974,
   Perturbation of a Slowly Rotating Black Hole by a Stationary Axisymmetric Ring of Matter.
   I. Equilibrium Configurations,
   ApJ 191, 521
\bibitem[Will(1975)]{bib:Will1975}
   Will C. M., 1975,
   Perturbation of a slowly rotating black hole by a stationary axisymmetric ring of matter.
   II - Penrose processes, circular orbits, and differential mass formulae,
   ApJ 196, 41
\bibitem[Zellerin \& Semer\'ak(2000)]{bib:ZellSem2000}
   Zellerin T. \& Semer\'ak O., 2000,
   Two-soliton stationary axisymmetric sprouts from Weyl seeds,
   Class. Quantum Grav. 17, 5103

\end{thebibliography}
\end{document}